\documentclass{article}[12pt] 
\usepackage{graphicx}
\usepackage{amsmath}
\usepackage{amssymb}
\usepackage{mathtools}
\usepackage[toc,page]{appendix}
\usepackage{hyperref}
\usepackage{amsfonts}
\usepackage{epstopdf}
\usepackage{overpic}
\usepackage{tikz}
\usepackage{float}
\usetikzlibrary{calc}
\usepackage{todonotes}
\def\Xint#1{\mathchoice
	{\XXint\displaystyle\textstyle{#1}}%
	{\XXint\textstyle\scriptstyle{#1}}%
	{\XXint\scriptstyle\scriptscriptstyle{#1}}%
	{\XXint\scriptscriptstyle\scriptscriptstyle{#1}}%
	\!\int}
\def\XXint#1#2#3{{\setbox0=\hbox{$#1{#2#3}{\int}$}
		\vcenter{\hbox{$#2#3$}}\kern-.5\wd0}}

\usepackage{color}  
\definecolor{qqwuqq}{rgb}{0,0.39215686274509803,0}
\definecolor{darkGreen}{rgb}{0,0.45,0}
\definecolor{darkBlue}{rgb}{0,0,0.7}
\definecolor{PineGreen}{rgb}{0,0.2,0}
\usepackage{hyperref}
\hypersetup{
	colorlinks=true, 
	linktoc=all,    
	linkcolor=darkBlue, 
	citecolor=darkGreen,
}
\usepackage{comment}

\usepackage[margin=3.6cm]{geometry}
\title{Large-amplitude membrane flutter in inviscid flow}
\author{Christiana~Mavroyiakoumou\footnote{chrismav@umich.edu}~~and   
      Silas~Alben\footnote{alben@umich.edu} \\Department of Mathematics, University of Michigan, \\Ann Arbor, MI 48109, USA.
}
\date{\today}

\begin{document}

\maketitle

\begin{abstract}
We study the large-amplitude flutter of membranes (of zero bending rigidity)  with vortex-sheet wakes in 2D inviscid fluid flows. We apply small initial deflections and track their exponential decay or growth and subsequent large-amplitude dynamics in the space of three dimensionless parameters: membrane pretension, mass density, and stretching modulus. With both ends fixed, all the membranes converge to steady deflected shapes with single humps that are nearly fore-aft symmetric, except when the deformations are unrealistically large. With leading edges fixed and trailing edges free, the membranes flutter with very small amplitudes and high spatial and temporal frequencies at small mass density. As mass density increases, the membranes transition to periodic and then increasingly aperiodic motions, and the amplitudes increase and spatial and temporal frequencies decrease. With both edges free, the membranes flutter similarly to the fixed-free case but also translate vertically with steady, periodic, or aperiodic trajectories, and with nonzero slopes that lead to small angles of attack with respect to the oncoming flow.
\end{abstract}


\section{Introduction}
\label{sec:intro}
There is a wealth of literature describing the fluid dynamics induced by the motion of a flexible body through a fluid flow. In most studies the body motion is bending-dominated, with a moderate
bending modulus, but essentially inextensible
 \cite{Taneda_JPhysSocJpn_1968,kornecki1976ait,ZCLS2000,ZP2002,watanabe2002esp,shelley2005heavy,AM2005,ESS2007,alben2008flapping,alben2008ffi,michelin2008vortex,eloy2008aic,shelley2011flapping}. A smaller number of studies, including this one, consider softer materials---extensible membranes---that have a negligible bending modulus, and undergo significant stretching in a fluid flow 
\cite{newman1991stability,lian2005numerical,sygulski2007stability,song2008aeromechanics,tiomkin2017stability,tzezana2019thrust}.
Examples are rubber, textile fabric, or the skin of swimming and flying animals \cite{newman1987aerodynamic,lian2003membrane,lauder2006locomotion,cheney2015wrinkle}.

Here we extend our model of an inextensible bending body with a vortex-sheet wake \cite{alben2008optimal,alben2009simulating} to an
extensible membrane. Our previous work investigated the nonlinear dynamics of a periodically pitching flexible body in a fluid stream~\cite{alben2009simulating} and the flapping-flag instability~\cite{alben2008flapping}, among many other studies of this problem~\cite{ESS2007,shelley2005heavy,tang2003flutter,michelin2008vortex,huang2010three,chen2014bifurcation,alben2015flag}.  As bending rigidity is decreased below the flutter threshold, the flag
transitions from periodic to chaotic dynamics. Our model includes the separation of vortex sheets at sharp edges~\cite{krasny1991vortex,nitsche1994numerical,jones2003separated,jones2005falling}, and regularizes free vortex sheets to avoid singularities~\cite{chorin1973discretization,krasny1986desingularization,brady1998regularized,nitsche2003comparison,alben2010regularizing}. 

In this model, vortex sheets approximate the thin viscous boundary layers along the body, which are advected from its trailing edge
into the flow downstream. This can be regarded as the inviscid limit of the viscous flow, and gives a good representation of the large-scale features of the flow and the vortex wake dynamics at Reynolds numbers of 
O($10^2$--$10^5$)
\cite{nitsche1994numerical,sheng2012simulating,xu2017computation}. These flows are challenging to simulate directly due to
the need to resolve sharp layers of vorticity in the vicinity of an unsteady, possibly deforming solid boundary. The immersed boundary method~\cite{ZP2002,peskin2002immersed,griffith2005order,taira2007immersed,ghias2007sharp,tytell2010interactions,hamlet2011numerical,tian2011efficient} can successfully simulate this class of problems. Very fine grids are needed to resolve the vorticity, and these are
refined adaptively for efficiency \cite{roma1999adaptive,griffith2007adaptive}. However, by only computing flow quantities on one-dimensional contours (the body and the vortex sheet wake), the vortex sheet model is orders of magnitude less expensive to compute.




The focus of the present study is extensible membrane
flutter: how a membrane initially aligned with a fluid flow becomes unstable to transverse deflections
and eventually reaches steady-state
large-amplitude dynamics. The initial, small-amplitude stage of the flutter
instability has been the focus of several experimental, theoretical, and numerical studies, summarized in table~\ref{table:summary}. We classify these and the present study in terms of three dimensionless parameters: membrane mass density, stretching modulus, and pretension. The present simulations allow us to consider wide ranges of the parameters including
those of the previous studies. Newman and Pa\"{i}doussis used
 an infinite periodic membrane model with a low-mode
 approximation and found that stability is lost through
 divergence \cite{newman1991stability}. Le Ma\^{i}tre {\it et al.}
 used a vortex sheet model to study a more complex situation---the motions of a sail membrane under harmonic perturbations of the trailing edge and with randomly perturbed inflow velocities~\cite{le1999unsteady}.
Sygulski studied the membrane flutter threshold and divergence modes theoretically, with some experimental validation \cite{sygulski2007stability}. Tiomkin and Raveh presented a more detailed flutter threshold calculation using an inviscid, small amplitude vortex sheet model~\cite{tiomkin2017stability}. Nardini {\it et al.} compared a reduced-order model with direct numerical
simulations to study the effect of Reynolds number on the flutter
stability threshold and small-amplitude membrane deflection modes~\cite{nardini2018reduced}. So far there has been relatively little work on the large-amplitude dynamics following the initial flutter instability, and
this is the focus of the present work.

\begin{table}[H]\label{table:summary}
\caption{Summary of parameter ranges used in previous and current membrane studies. Computational ($^c$), experimental ($^e$), or theoretical ($^t$) ranges of the dimensionless body mass density $R_1$, stretching modulus $R_3$, and pretension $T_0$, defined in section \ref{nondim}. \label{tab:tableCrit}}
\centering
\vspace{.1cm}
\begin{tabular}{lccc}
\hline\noalign{\smallskip}
Study & $R_1=\displaystyle\frac{\rho_s h}{\rho_f L}$ & $R_3 = \displaystyle\frac{Eh}{\rho_f U^2L}$ & $T_0=\displaystyle\frac{\overline{T}}{\rho_f U^2L W}$\\
\noalign{\smallskip}\hline\noalign{\smallskip}
Newman \textit{et al.} (1991)$^t$  & 0--6 & --- & 0--2\\
Le Ma\^\i tre  \textit{et al.} (1999)$^e$   & 0--0.8 & 10,\,50,\,100,\,500,\,1000 & --- \\
Sygulski (2007)$^{e\,\&\,t}$  & 0.1,\,1 & --- & 130.6,\,217 \\
Jaworski \textit{et al.} (2012)$^{c \,\& \,e}$   & 1.2 & 100,\,200,\,400,\,614 & 4,\,10,\,20,\,30.7  \\
Tiomkin \textit{et al.} (2017)$^c$ & 0--80 & --- & 0--6 \\
Nardini \textit{et al.} (2018)$^c$ & 0--60 & --- & 0--3 \\
Present study$^{c\,\&\,t}$   & $0.001$--$100$ & $1$--$10000$ &$0.001$--$1000$ \\
\noalign{\smallskip}\hline
\end{tabular}
\end{table}

Another main topic of fluid-membrane interaction studies is bio-inspired propulsion. Piquee {\it et al.} studied how a membrane wing adjusts its shape to fluid pressure loading at various angles of attack~\cite{piquee2018aerodynamic}. Tzezana and Breuer coupled thin airfoil theory with a membrane equation to study the effects of wing compliance, inertia, and flapping kinematics on aerodynamic performance~\cite{tzezana2019thrust}. Jaworski and Gordnier studied a heaving and pitching membrane airfoil in a fluid stream numerically at Reynolds number 2500, and found elastic modulus and prestress parameters that led to enhanced thrust and propulsive efficiency~\cite{jaworski2012high}. The passive adaptivity of a membrane wing has the potential to increase lift forces and delay the occurrence of stall to higher angles of attack for micro-air vehicles (MAVs)~\cite{hu2008flexible,lian2005numerical,stanford2008fixed}.
Schomberg {\it et al.} used electrostatic
forces to control a membrane shape and delay the transition from a laminar boundary layer, reducing viscous drag~\cite{schomberg2018transition}. Here, however, we focus solely on the flutter problem and defer propulsion-related dynamics to future work.


\section{Membrane-vortex-sheet model}\label{sec:membrane model}
\begin{figure}[H]
	\centering
	\includegraphics[width=.75\linewidth]{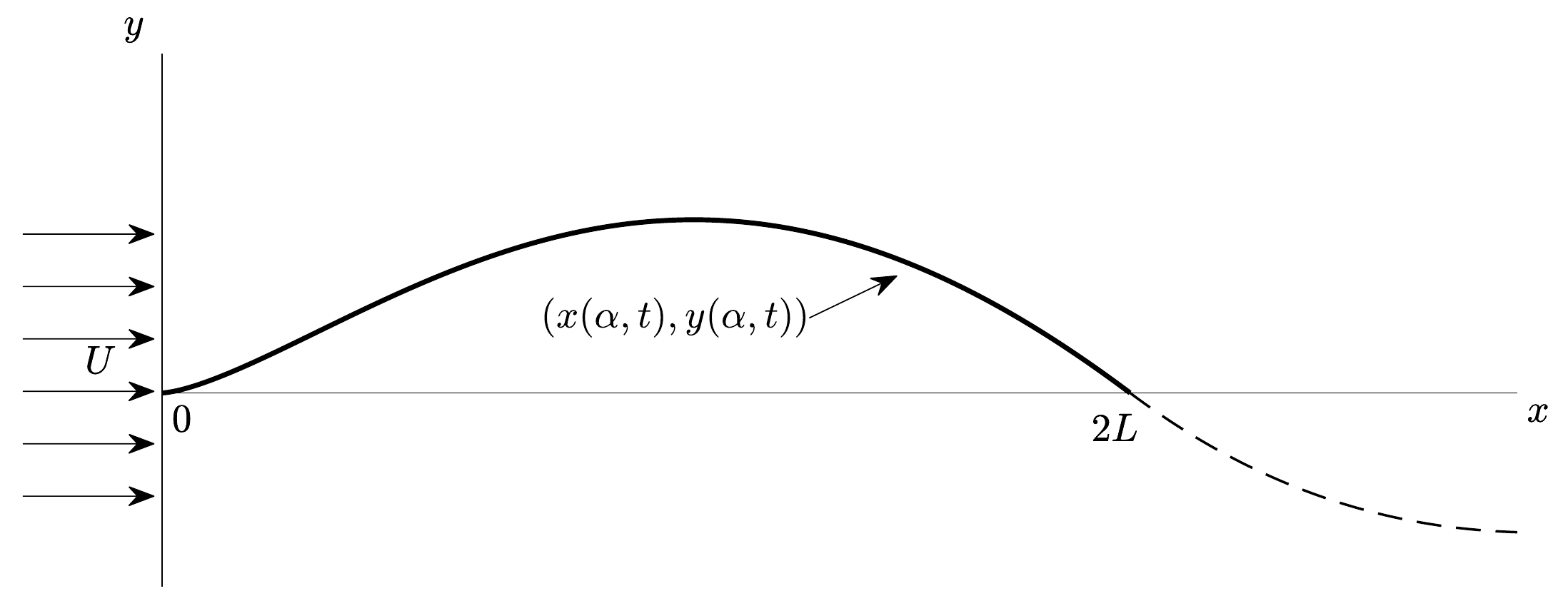}
	\caption{Schematic diagram of a flexible membrane (solid curved line) at an instant in time. Here $2L$ is the chord length (the distance
	between the endpoints), $U$ is the oncoming flow velocity, $(x(\alpha,t),y(\alpha,t))$ is the membrane position, and the dashed line is the free vortex sheet wake.}\label{fig:schematic}
\end{figure}

We first consider the motion of an extensible membrane that is fixed at two endpoints and held in a two-dimensional fluid flow, like much of the previous work. A uniform background flow is prescribed with velocity $U$, directed parallel to the chord connecting the endpoints (see figure~\ref{fig:schematic}). The instantaneous position of the membrane is given by $\mathbf{X}(\alpha,t)=(x(\alpha,t),y(\alpha,t))$, parameterized by the material coordinate $\alpha, -L\leq \alpha\leq L$ ($L$ is half the chord length), and time $t$. It is convenient to also describe the membrane position in complex notation, $\zeta(\alpha,t)=x(\alpha,t)+iy(\alpha,t)$. The inviscid flow can be represented by a vortex sheet---a curve across which the tangential velocity component is discontinuous~\cite{saffman1992vortex}---and whose position
and strength evolve in time. The vortex sheet consists of two parts. One is ``bound" (it coincides with the membrane, for $-L\leq \alpha\leq L$), and the other is ``free," emanating from the trailing edge of the membrane at $\alpha=L$. We parameterize the free sheet by arc length $s$ or by a material fluid coordinate $\Gamma$. The bound and free vortex sheets have strength densities denoted by $\gamma$ and positions denoted by $\zeta$.

The membrane dynamics are described by the unsteady extensible elastica equation with body inertia, stretching resistance, and fluid pressure loading~\cite{tadjbakhsh1966variational}: 
\begin{equation}
\rho_s h W \partial_{tt}\zeta(\alpha,t) =\partial_\alpha(T(\alpha,t)\hat{\mathbf{s}})-[p](\alpha,t)W\partial_\alpha s\hat{\mathbf{n}}, \label{beam}
\end{equation}
where $\rho_s$ is the mass per unit volume of the
membrane,
$h$ is its thickness,
and $W$ is its out-of-plane width, all uniform along the length in the undeformed state. The material coordinate $\alpha\in[-L,L]$ is 
the $x$-coordinate of the membrane in the initial flat, uniformly prestretched state.
Other quantities that appear in (\ref{beam}) are $\hat{\mathbf{s}},\hat{\mathbf{n}}\in \mathbb{C}$, which represent the unit vectors tangent and normal to the membrane, respectively,
\begin{equation}
\hat{\mathbf{s}}=\partial_\alpha\zeta(\alpha,t)/\partial_\alpha s(\alpha,t)=e^{i\theta(\alpha,t)}\quad\text{and}\quad \hat{\mathbf{n}}=i\hat{\mathbf{s}}=ie^{i\theta(\alpha,t)},
\end{equation}
with $\theta(\alpha,t)$ the local tangent angle,
$s(\alpha,t)$ the local arc length coordinate,
and $\kappa(\alpha,t)=\partial_\alpha\theta/\partial_\alpha s$ the membrane's curvature.
$[p](\alpha, t)$ is the local pressure difference
across the membrane, from the side
toward which $\hat{\mathbf{n}}$ points to the other side. 

The membrane tension $T(\alpha,t)$ is given by linear elasticity \cite{carrier1945non,carrier1949note,narasimha1968non,nayfeh2008linear}:
\begin{equation}
T=\overline{T}+EhW(\partial_{\alpha}s-1), \label{T}
\end{equation}
where $E$ is the Young's modulus. Thus the tension is a constant $\overline{T}$, the  
``pretension," in the (initial) undeflected equilibrium state $\zeta(\alpha,0)=\alpha$.

The local stretching
factor is $\partial_{\alpha}s = \sqrt{\partial_\alpha x^2 + \partial_\alpha y^2} \approx 1 + \partial_\alpha y^2/2$ for small deflections ($x \approx \alpha, \partial_\alpha y \ll 1$), so by (\ref{T}) $T$ is
a constant plus a term
that is quadratic in deflection for small deflections. The normal
component of the tension force (the first term
on the right side of (\ref{beam})) is $T\kappa\partial_\alpha s$, and thus has
terms that are linear (proportional to $\overline{T}$) and cubic (proportional to $EhW$) in deflection for small deflections.

We note that equation (\ref{beam}) is obtained by writing a force balance equation for a small section of membrane lying between $\alpha$
and $\alpha + \Delta\alpha$:
\begin{equation}
\rho_s h W \partial_{tt}\zeta(\alpha,t)\Delta\alpha =T(\alpha+\Delta\alpha,t)\hat{\mathbf{s}}-T(\alpha,t)\hat{\mathbf{s}} -[p](\alpha,t)\hat{\mathbf{n}}W\left(s(\alpha+\Delta\alpha,t)-s(\alpha,t)\right). \label{beam0}
\end{equation}
Dividing by $\Delta\alpha$ and taking the limit 
$\Delta\alpha \to 0$, we obtain (\ref{beam}).



\subsection{Nondimensionalization \label{nondim}}

We nondimensionalize the governing equations by the density of the fluid $\rho_f$, the half-chord $L$, and the imposed fluid flow velocity $U$. The membrane equation (\ref{beam}) becomes 
\begin{equation}\label{dim1}
\frac{\rho_s h U^2W}{L}\widetilde{\partial_{tt}\zeta}=\frac{1}{L} \widetilde{{\partial_\alpha}}[(\overline{T}+EhW\widetilde{(\partial_{\alpha}s-1))\hat{\mathbf{s}}]}-\rho_fU^2W\widetilde{[p]\partial_\alpha s\hat{\mathbf{n}}},
\end{equation}
with dimensionless quantities denoted by tildes. Dividing (\ref{dim1}) by $\rho_f U^2 W$ throughout yields
\begin{equation}
\frac{\rho_s h }{\rho_f L}\widetilde{\partial_{tt}\zeta}=-\widetilde{[p]\partial_\alpha s\hat{\mathbf{n}}}+\frac{1}{\rho_fU^2LW}\widetilde{{\partial_\alpha}}[(\overline{T}+EhW\widetilde{(\partial_{\alpha}s-1))\hat{\mathbf{s}}]}.
\end{equation}
Thus the dimensionless membrane equation (dropping tildes) is
\begin{equation}\label{eq:mainext}
R_1\partial_{tt}\zeta-\partial_\alpha\left((T_0+R_3(\partial_{\alpha}s-1))\hat{\mathbf{s}}\right)=-[p]\partial_\alpha s\hat{\mathbf{n}}.
\end{equation}
The dimensionless parameters of the membrane are
\begin{equation}
R_1=\frac{\rho_s h}{\rho_f L}, \,\,\,  T_0=\displaystyle\frac{\overline{T}}{\rho_f U^2L W},\,\,\text{and}\,\,\, R_3=\frac{Eh}{\rho_f U^2 L},
\end{equation}
where $R_1$ is the dimensionless membrane mass density, $T_0$ is the dimensionless pretension, and $R_3$ is
the dimensionless stretching modulus. 
We assume that the aspect ratio $h/L$ is small, but $\rho_s/\rho_f$ may be large, so $R_1$ may assume any non-negative value. We have neglected
bending rigidity, denoted $R_2$ in \cite{alben2008flapping}.
In the extensible membrane regime, $R_3$ is finite, so $R_2 = R_3 h^2 / 12 L^2 \to 0$ in the limit $h/L\to 0$. We have also neglected the
effects of
rotary inertia and the Poisson ratio (the transverse contraction due to axial stretching),
which have usually been neglected at leading order
in nonlinear membrane models \cite{nayfeh2008linear} and
the aforementioned membrane studies. The rotary inertia and
bending rigidity terms are given in
\cite{tadjbakhsh1966variational}, and
those involving Poisson ratio are given
in \cite{nayfeh2008linear}.


For large-Reynolds-number flows, there are thin viscous boundary layers along the sides of the membrane. Across these boundary layers, the component of fluid velocity that is tangent to the membrane is brought to zero on the membrane~\cite{batchelor1967introduction}. When the fluid in the boundary layer is advected off of the membrane's trailing edge, a free shear layer forms~\cite{saffman1992vortex,alben2010regularizing}. In the limit of large Reynolds number, the
two boundary layers tend to vortex
sheets which coincide as a single bound vortex sheet (approximating the body thickness as zero for
the fluid computation). The free shear layer tends to a free vortex sheet~\cite{saffman1992vortex}. 
The free sheet circulation is defined as an integral of the vortex sheet strength $\gamma$ (the jump in the tangential component of the fluid velocity) over the free vortex sheet
\begin{equation}\label{eq:gammaGamma}
\Gamma(s,t)=-\int_{s}^{s_{\text{max}}}\gamma(s',t)\text{d}s',\quad 0<s<s_{\text{max}},
\end{equation}
where $s$ is arc length along the free sheet, starting
from 0 where the free sheet meets the membrane's trailing edge and 
ending at $s_{\text{max}}$ at the free sheet's far end. 
Following~\cite{jones2003separated,alben2009simulating} we define
the (negative of the) total circulation in the free sheet:
\begin{equation}
    \Gamma_{+}(t)=\Gamma(0,t)=-\int_{0}^{s_{\max}}\gamma(s',t)\,\text{d}s'.
\end{equation}
The complex conjugate of the flow velocity $\mathbf{u} = (u_x,u_y)$ at any point $z$ in the flow (not on the vortex sheets) can be calculated in terms of $\gamma$ by integrating the vorticity in the bound and free vortex sheets against the Biot-Savart kernel~\cite{saffman1992vortex}
\begin{equation}\label{eq:biotSavart}
u_x(z)-iu_y(z)=1 + \frac{1}{2\pi i}\int_{-1}^{1}\frac{\gamma(\alpha,t)}{z-\zeta(\alpha,t)}\partial_\alpha s(\alpha,t)\,\text{d}\alpha
+ \frac{1}{2\pi i}\int_{0}^{s_{\max}}\frac{\gamma(s,t)}{z-\zeta(s,t)}\,\text{d}s
,
\end{equation}
with unity on the right
hand side representing
the imposed background flow and
the dimensionless material coordinate $\alpha$ ranging from $-1$ to 1 on the membrane.
By Kelvin's circulation theorem,
$\Gamma$ is conserved at fixed material elements of the free vortex sheet.
We reparameterize the free sheet position as $\zeta(\Gamma,t)$ and evolve the
position at a fixed $\Gamma$ simply
by following the local fluid velocity. 
This is done by taking the average
of the limits of 
(\ref{eq:biotSavart}) as $z$ approaches $\zeta(\Gamma,t)$ from both sides~\cite{saffman1992vortex}:
\begin{equation}\label{eq:birkhoffrott}
\frac{\partial\overline{\zeta}}{\partial t}(\Gamma,t)=1 + \frac{1}{2\pi i}\int_{-1}^1\frac{\gamma(\alpha,t)}{\zeta(\Gamma,t)-\zeta(\alpha,t)}\partial_\alpha s(\alpha,t)\,\text{d}\alpha-\frac{1}{2\pi i }\Xint -_{0}^{\Gamma_+(t)}\frac{\text{d}\Gamma'}{\zeta(\Gamma,t)-\zeta(\Gamma',t)}.
\end{equation}
In (\ref{eq:birkhoffrott}), $\partial\overline{\zeta}/\partial t$ is the complex conjugate velocity
at $\zeta(\Gamma,t)$, and the second integral is a Cauchy-principal-value integral. We have reparameterized the free sheet integral using $\gamma\partial_\alpha s\,\text{d}\alpha=-\text{d}\Gamma$.

We may solve for the bound vortex sheet strength $\gamma(\alpha, t)$ in terms
of the membrane velocity by
equating the components of the fluid
and membrane velocities normal to the membrane (``the kinematic condition"), which are found by taking
the average of the limits of 
(\ref{eq:biotSavart}) as $z$ approaches
the {\it membrane} from both sides:
\begin{align}
\text{Re}(\hat{\mathbf{n}}\,\partial_t\overline{\zeta}(\alpha,t))&=\text{Re}\left(\hat{\mathbf{n}}\left( 1+\frac{1}{2\pi i} \Xint-_{-1}^{1}\frac{\gamma(\alpha',t)\partial_\alpha s(\alpha',t)}{\zeta(\alpha,t)-\zeta(\alpha',t)}\,\text{d}\alpha'-\frac{1}{2\pi i }\int_{0}^{\Gamma_+(t)}\frac{\text{d}\Gamma'}{\zeta(\alpha,t)-\zeta(\Gamma',t)}\right)\right).\label{kinematic}
\end{align}




When the left hand side and the second integral on the right hand side of~\eqref{kinematic} are known, the general solution $\gamma(\alpha,t)$ has inverse-square-root singularities at $\alpha=\pm 1$. Therefore we define $v(\alpha,t)$, the bounded part of $\gamma(\alpha,t)$, by 
\begin{equation}\label{strength}
\gamma(\alpha,t)=\frac{v(\alpha,t)}{\sqrt{1-\alpha^2}}.
\end{equation}
An additional scalar constraint is required to uniquely specify the solution $\gamma$ (or $v$) to~\eqref{kinematic}.
It is the conservation of total circulation (Kelvin's circulation theorem):
\begin{equation}
\int_{-1}^{1}\gamma\partial_\alpha s\,\text{d}\alpha = \int_{-1}^{1}\frac{v(\alpha,t)}{\sqrt{1-\alpha^2}}\partial_{\alpha}s\,\text{d}\alpha = \Gamma_+(t).
\end{equation}


In (\ref{eq:birkhoffrott}) and (\ref{kinematic}) it is
helpful to replace the free-sheet integral with a regularized version to avoid singularities in the sheet curvature  \cite{krasny1986desingularization}.
The second integral in (\ref{eq:birkhoffrott}) becomes
\begin{equation}\label{eq:bsmoothed}
-\frac{1}{2\pi i}\Xint -_{0}^{\Gamma_+(t)}\frac{\overline{\zeta(\Gamma,t)-\zeta(\Gamma',t)}}{|\zeta(\Gamma,t)-\zeta(\Gamma',t)|^2+\delta(\Gamma',t)^2}\,\text{d}\Gamma',
\end{equation}
with a regularization parameter
\begin{equation}
   \delta(\Gamma,t)=\delta_0\left(1-e^{-s(\Gamma,t)^2/\varepsilon^2}\right). 
\end{equation}
The effect of $\delta$ is to inhibit the growth of free sheet structures
(e.g.\ inner turns of spirals) on scales smaller than $\delta$ while maintaining the shape and motion of the sheet on larger scales. Our choice of $\delta$ tends to 0 quadratically over a scale given by $\varepsilon$ as the membrane trailing edge is approached, to decrease the effect of regularization on the flow near the trailing edge and the production of circulation \cite{alben2009simulating,alben2010regularizing}.
Here we set $\varepsilon$ to 0.4 and $\delta_0$ to 0.2, choices that make the effect of regularization on circulation production small without a significant increase in the total number of points
needed to resolve the free sheet 
\cite{alben2010regularizing}. The Kutta condition determines the rate of
circulation production $\text{d}\Gamma_+(t)/\text{d}t$ by making
the fluid velocity at the trailing edge
finite. This means $\gamma(1,t)$ must be finite, and thus $v(1,t)=0$ by~(\ref{strength}).


The vortex sheet strength $\gamma(\alpha,t)$ is coupled to the pressure jump $[p](\alpha,t)$ across the membrane using a version of the unsteady Bernoulli equation written
at a fixed material point on the membrane:
\begin{equation}\label{eq:pressureAlpha}
\partial_\alpha s\partial_t\gamma+(\mu - \tau)\partial_\alpha\gamma +\gamma(\partial_\alpha\mu-\partial_\alpha s\nu \kappa)=\partial_\alpha[p].
\end{equation}
This equation is derived in appendix~\ref{app:pressureJump},
and generalizes the derivation in \cite[appendix A]{alben2012attraction} to the case of an extensible body.

In (\ref{eq:pressureAlpha}), $\mu$ is the tangential component of the average flow velocity at the membrane,
\begin{equation}\label{eq:mu}
\mu(\alpha,t)=\text{Re}\left(\overline{\hat{\mathbf{s}}}\left( 1+\frac{1}{2\pi i} \Xint-_{-1}^{1}\frac{\gamma(\alpha',t)\partial_\alpha s(\alpha')}{\zeta(\alpha,t)-\zeta(\alpha',t)}\,\text{d}\alpha'-\frac{1}{2\pi i }\int_{0}^{\Gamma_+(t)}\frac{\text{d}\Gamma'}{\zeta(\alpha,t)-\zeta(\Gamma',t)}\right)\right),
\end{equation}
and $\tau$ and $\nu$ are the components of the membrane's velocity tangent and normal to itself,
respectively:
\begin{equation}\label{eq:tau}
\tau(\alpha,t)=\text{Re}\left(\partial_t\zeta(\alpha,t)\overline{\hat{\mathbf{s}}}\right) \quad ; \quad \nu(\alpha,t)=\text{Re}\left(\partial_t\zeta(\alpha,t)\overline{\hat{\mathbf{n}}}\right).
\end{equation}
The pressure jump across the free sheet is zero, which yields
\begin{equation}
[p]|_{\alpha=1}=0,
\end{equation}
the boundary condition we use to integrate \eqref{eq:pressureAlpha} and obtain
$[p](\alpha,t)$ on the membrane.


\subsection{Boundary and initial conditions}
\begin{figure}[H]
	\includegraphics[width=\textwidth]{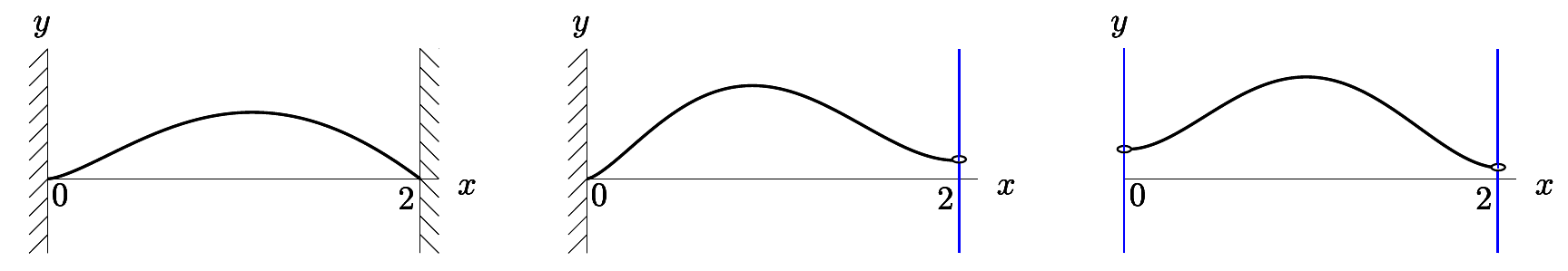}	\caption{Schematic diagrams of the three sets of boundary conditions considered: fixed-fixed (left), fixed-free (center), and free-free (right).}\label{fig:bc}
\end{figure}
We investigate three cases of boundary conditions at the two ends of the membrane, shown schematically in figure~\ref{fig:bc}. In all three cases, the $x$-coordinates of the ends are constant: $x(-1,t)=0$ and $x(1,t)=2$.
In the first case, ``fixed-fixed," the membrane is flat at $t=0$, and we set the deflection to zero at both ends of the membrane after a small initial perturbation.
More precisely, we smoothly perturb $y$ at the leading edge slightly away from zero and relax it to zero exponentially in time:
\begin{equation}
    y(-1,t) = \sigma \left(\frac{t}{\eta}\right)^3e^{-(t/\eta)^3}\label{pert},
\end{equation} 
where $\sigma$ is a constant chosen in the range $10^{-6}$--$10^{-3}$ (depending on whether small or large amplitude dynamics are studied) and $\eta=0.2$. We set the trailing edge deflection $y(1,t)$ to zero for all~$t$.
This is essentially the case considered by most previous studies of membrane flutter~\cite{le1999unsteady,sygulski2007stability,tiomkin2017stability,nardini2018reduced}, and here we find, surprisingly, that all physically-reasonable deflected membrane states are steady (i.e.\ without oscillations). In the second case, ``fixed-free," we again make the membrane flat initially and then
set the leading edge position according to~(\ref{pert}), but allow the trailing edge to deflect freely in the vertical direction. This is the classical free-end boundary condition for a membrane~\cite{graff1975wave,farlow1993partial} and corresponds to the membrane end fixed to a massless ring that slides without friction along a vertical pole (see figure \ref{fig:bc}, center). Without friction, the force from the pole on the ring at the membrane end is horizontal. The tension force from the membrane on the ring must also be horizontal,
or else the ring would have an infinite vertical acceleration. 
Therefore, $\hat{\mathbf{s}} = \hat{\mathbf{e}}_x$ at the
trailing edge, or equivalently, $\partial_\alpha y(1,t)=0$. 

Although well-known in classical mechanics, free-end boundary conditions have not been studied much in membrane flutter problems. We will show, however, that
free ends allow for a wide range of unsteady membrane dynamics, unlike in the fixed-fixed case. Related work has studied the dynamics and flutter of membranes and cables under gravity with free ends~\cite{triantafyllou1994dynamic,manela2017hanging}. Here we neglect gravity to focus specifically on the basic flutter problem~\cite{shelley2011flapping}. Without gravity, some restriction on the motion of the free membrane ends is needed to avoid
ill-posedness due to membrane compression~\cite{triantafyllou1994dynamic}. This is provided by the vertical poles in figure \ref{fig:bc}.  Although this type of free-end boundary condition has mainly been studied theoretically, it has been realized experimentally by \cite{kashy1997transverse},
with the membrane represented by an extensional spring that is tethered by steel wires to vertical supports. Membrane flutter with free ends may relate to other cases in which the membrane ends are only partially constrained, such as energy harvesting
by membranes in tensegrity structures
\cite{sunny2014optimal,yang2016modeling}. 

In the third case, ``free-free," both ends are free: $\partial_{\alpha}y(-1,t)=\partial_{\alpha}y(1,t)=0$. Here the membrane is perturbed differently: it is initially set with a small nonzero
slope, $\zeta(\alpha,0) = (\alpha+1)(1+i\sigma)$
for $\sigma=10^{-6}$--$10^{-3}$ (again depending on whether small or large amplitude dynamics are studied).

We investigate the dynamics for the different sets of boundary conditions in detail in the sections that follow.



\section{Numerical results and discussion}\label{sec:simulations}

We now describe the range of dynamics of a two-dimensional extensible membrane with the three sets of boundary
conditions. In each case, we first present the flutter stability region for the flat membrane in the $R_1$-$T_0$ plane (it is independent of $R_3$ because it depends only on the small-deflection behavior). We then consider the large-amplitude dynamics using three main quantities to characterize them. One is the time-averaged deflection of the membrane, 
\begin{align}
\langle y_{\mbox{defl}} \rangle \equiv  \frac{1}{t_2}\int_{t_1}^{t_1+t_2} \left(\max_{-1\leq\alpha\leq 1} y(\alpha,t) - \min_{-1\leq\alpha\leq 1} y(\alpha,t)\right) \text{d}t, \label{ydefl}
\end{align}
where $t_1$ and $t_2$ are sufficiently large (typically 50--100) that $\langle y_{\mbox{defl}} \rangle$ changes by less than 1\% with further increases in these values. $\langle y_{\mbox{defl}} \rangle$ is the maximum membrane deflection minus the minimum deflection, averaged over time. 

The second quantity is the frequency, defined as the mean frequency in the power spectrum computed using Welch's method~\cite{welch1967use}. The power spectrum is obtained from a time series of the free sheet circulation when the membrane has reached steady-state large-amplitude dynamics.
The third quantity is the time-averaged number of zero crossings along the membrane, computed with the same temporal data as the power spectrum. The number of
zero crossings is a measure of the ``waviness" of the membrane shape.



\subsection{Fixed-fixed membranes \label{FixedFixed}}

We begin by presenting the dynamics of membranes with both ends fixed, the case considered by previous studies on membrane flutter. The most detailed linear stability analysis of the problem is by Tiomkin and Raveh \cite{tiomkin2017stability}. Their model is essentially a linearized version of ours, and includes a flat vortex wake extending to infinity downstream. They find
that all membranes become unstable when
the pretension $T_0$ drops below a critical value $\approx 1.73$, independent of $R_1$. A qualitatively similar result was found by \cite{newman1991stability} for an infinite
periodic membrane with no free vortex wake.
Below the critical pretension, the membranes in \cite{tiomkin2017stability} lose stability by divergence (exponential growth of deflection) at small $R_1$ or by divergence with flutter (exponential growth
with a complex growth rate--i.e.\ growth with oscillation) at
large $R_1$.

\begin{figure}[H]
\centering
\begin{overpic}[width=.95\textwidth,tics=10]{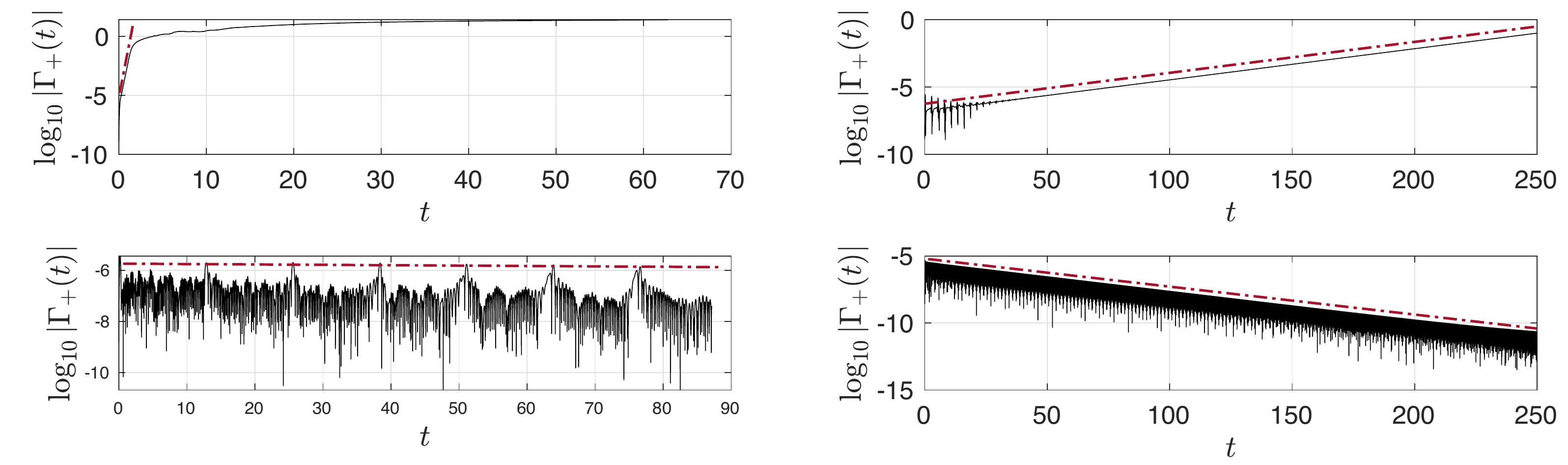}
	\put(-4,28){(a)}
	\put(48,28){(b)}
	\put(-4,13){(c)}
	\put(48,13){(d)}
\end{overpic}
\caption{Examples of the log$_{10}$ of the total wake circulation versus time with (a) $(R_1,T_0)=(10^{-1.8},10^{-1.2})$, (b) $(R_1,T_0)=(10^{-0.2},10^{0.2})$, (c) $(R_1,T_0)=(10^{2},10^{1})$, and (d) $(R_1,T_0)=(10^{1},10^{3})$. The growth/decay rates are given by the slopes of the dot-dashed red lines. In all cases we set $R_3 = 1$, but the linear growth rates are independent of $R_3$ (which multiplies a nonlinear term, negligible at small deflections).}\label{fig:examplesOfGamma}
\end{figure}
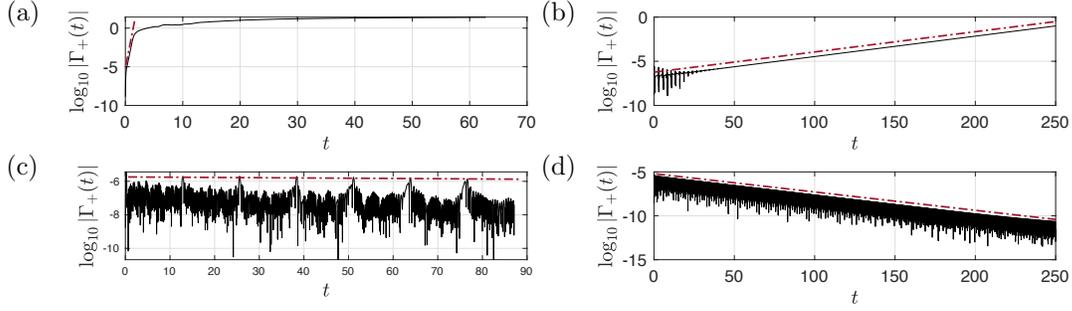

We use our nonlinear simulation to compute the stability threshold for membranes by applying the small transient perturbation (\ref{pert}) 
at the leading edge and observing exponential growth (followed by large-amplitude, nonlinear dynamics) or exponential decay, in membrane deflection and wake circulation $\Gamma_+(t)$. Examples are shown in figure \ref{fig:examplesOfGamma}.





\begin{figure}[H]
	\centering
	\begin{overpic}[width=.64\textwidth,tics=10]{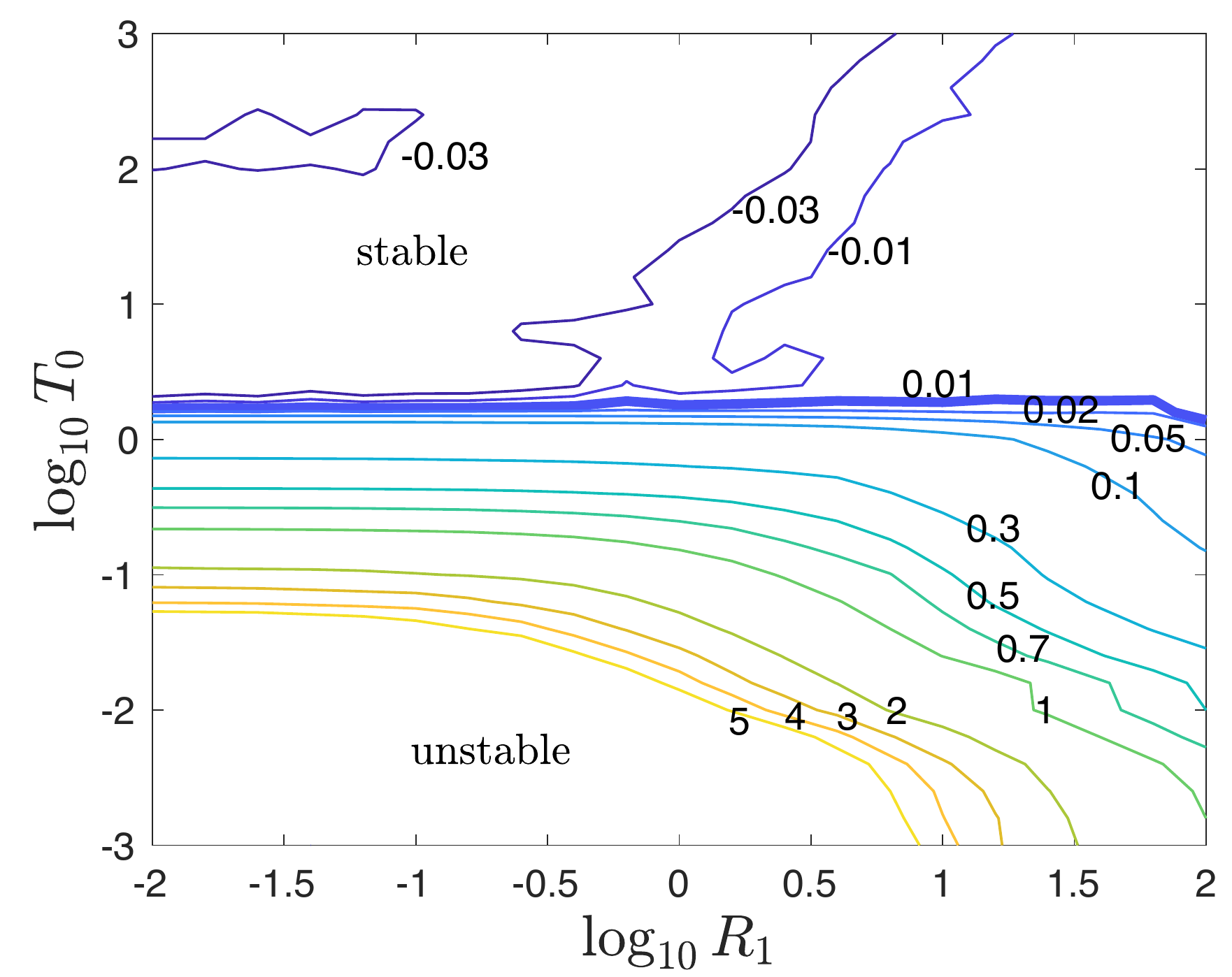}
	\end{overpic}
	\caption{A contour plot of the exponential (base 10) growth and decay rates of wake circulation after a small transient perturbation, in the fixed-fixed case. The thicker line separates the stable and unstable cases.}\label{fig:stability}
\end{figure}


We compute the growth and decay rates of the initial perturbation from data analogous to those in figure~\ref{fig:examplesOfGamma}, on a fine grid of values in the $R_1$-$T_0$ space spanning several orders
of magnitude in each parameter. Figure~\ref{fig:stability} is a contour plot of the growth/decay rates, i.e.\ $\beta$ in the early-time interval where $\Gamma_+(t) \approx K \,10^{\beta t}$ for some constant $K$. Values (well) above 5 occur in the lower left corner but are omitted for visual clarity. Above a critical pretension $T_0 \approx 1.78$
the membranes are stable, with small transient deflections decaying to the
flat state. Below the critical pretension we have a divergence instability: small transient deflections grow exponentially at
rate that is purely real. Tiomkin's critical pretension (in a slightly different model) is 1.73 \cite{tiomkin2017stability}; unlike that study, we do not find evidence of neutral flutter or divergence with flutter in the fixed-fixed case at any $R_1$. The main differences are that our wake length grows from zero while
that in \cite{tiomkin2017stability} is infinite, and our model
is a nonlinear, unsteady version of that in~\cite{tiomkin2017stability}. 

\begin{figure}[H]
	\centering
	\begin{overpic}[width=.9\textwidth,tics=10]{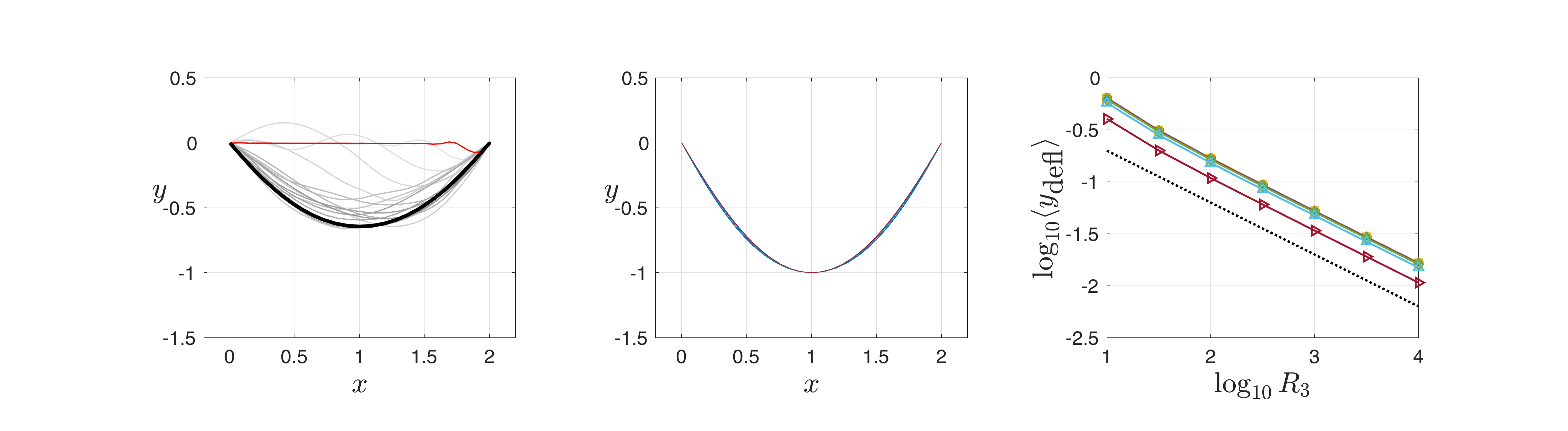}
		\put(-3,25){(a)}
		\put(32,25){(b)}
		\put(65,25){(c)}
	\end{overpic}
	\caption{Fixed-fixed membrane shapes. (a) For $R_1=10^{-0.5}$, $R_3=10^{1}$, and $T_0=10^{-3}$ the membrane shapes at $t$ = 1 (red), 1.5, 2.5, 3.5, 4.5, \ldots, 16.5 (light gray to black), (b) steady membrane shapes at late times, with $y$ coordinates scaled by maximum deflection, for $R_3$ ranging from $10^{1}$ to $10^4$ and $T_0$ ranging from $10^{-3}$ to 1. (c) The maximum membrane deflection at steady state versus the stretching modulus $R_3$.}\label{fig:amplitudeFixedsubplots}
\end{figure}

During the initial stages of the divergence instability, 
the membrane deflection grows from small amplitude without change of shape. Nonlinearities become important when the amplitude reaches
order one, and the membrane shape evolves to its eventual steady state. Figure~\ref{fig:amplitudeFixedsubplots}(a) shows a sequence of membrane snapshots during the nonlinear dynamics. The earliest
shape (red) is similar to those during the linear instability, with largest deflection near the trailing edge. Subsequent shapes
(ranging from light to dark gray and black) 
show the evolution to the eventual
steady-state. The final membrane shape
is nearly fore-aft symmetrical, similar to those
in \cite{waldman2017camber,nardini2018reduced,tzezana2019thrust}
but with larger deflection at this moderate choice of $R_3$.

\begin{figure}[H]
	\centering
	\begin{overpic}[width=.8\textwidth,tics=10]{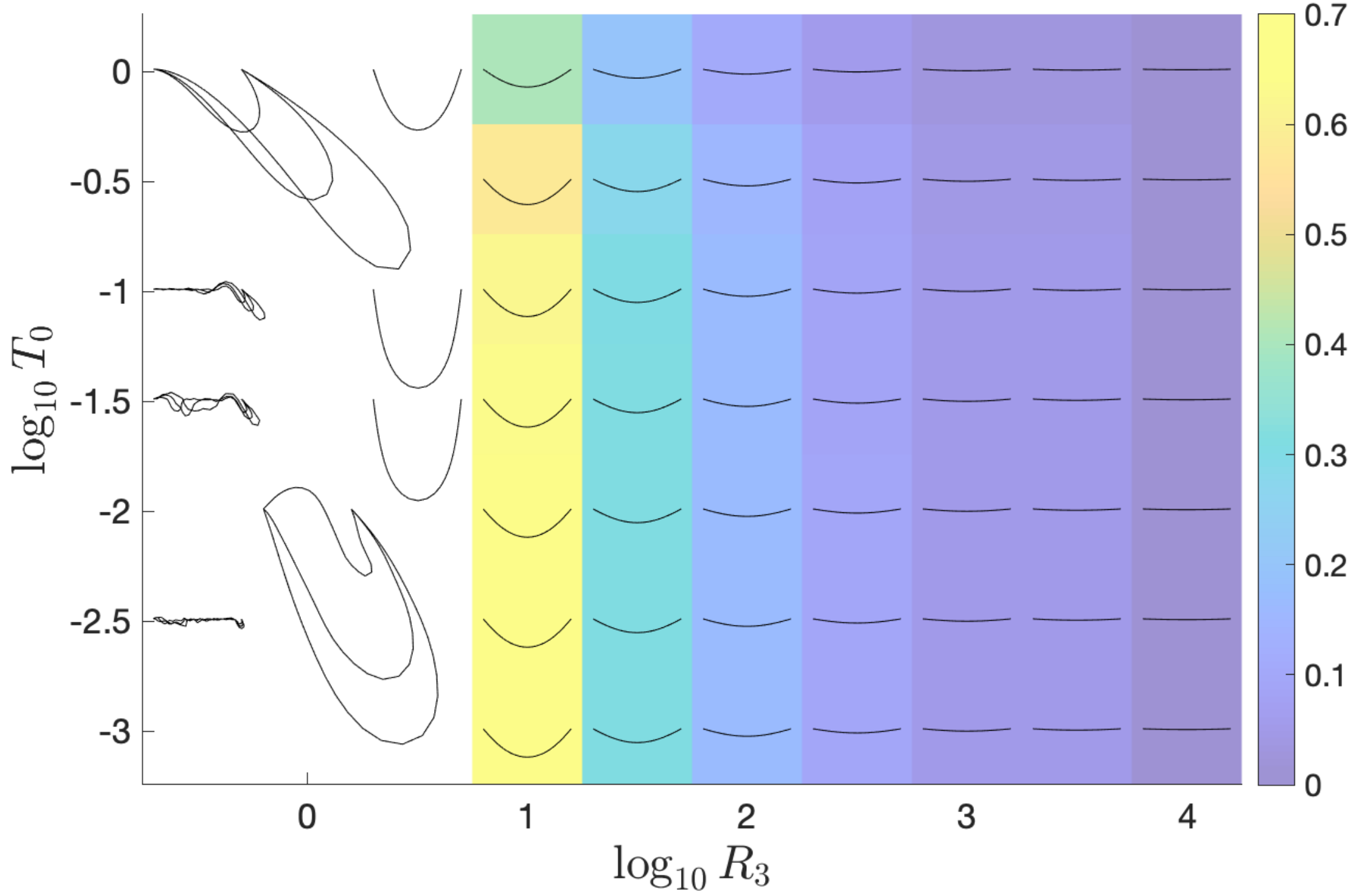}
	\end{overpic}
	\caption{Membrane profiles in the fixed-fixed case, at steady state with moderate deflections (colored background), or unphysically large or unsteady deflections (white background). In the unsteady cases a few snapshots at large times are shown. The colors show the deflection of the membrane (equation (\ref{ydefl})). Here, $R_1=10^{-0.5}$ but the steady shapes are independent of $R_1$.}\label{fig:amplitudeFixed}
\end{figure}

Figure~\ref{fig:amplitudeFixed} shows the late-time membrane shapes
across several decades of $R_3$ and $T_0$. The shapes are all steady for $R_3 > 10^{0.5}$ (and thus independent of $R_1$ here, as the acceleration term in~(\ref{eq:mainext}) is zero), but may oscillate chaotically for $R_3 \leq 10^{0.5}$, in which case a few snapshots are shown at late times. In these cases, the deflections are so large as to violate the assumption
that vortex shedding is confined to the trailing edge, so we do not consider them further. For $R_3 > 10^{0.5}$, the colors show the maximum membrane deflection. At a given $R_3$, the deflection increases slightly as $T_0$ decreases from the stability threshold to $\approx 0.1$, then converges as $T_0$ decreases
further, the $T_0$ term becoming insignificant in the membrane equation (\ref{eq:mainext}). At a given $T_0$, the
deflection decreases with increasing $R_3$ as a power law. Figure~\ref{fig:amplitudeFixedsubplots}(b) shows that the shapes are almost identical, however, when the amplitudes are normalized. Figure~\ref{fig:amplitudeFixedsubplots}(c) shows that the deflection 
$\langle y_{\mbox{defl}}\rangle \sim 1/\sqrt{R_3}$ (for small $T_0$ equal to $10^{-2}$ here).

We explain how the scaling $\langle y_{\mbox{defl}}\rangle\sim 1/\sqrt{R_3}$ arises from the $y$-component of the membrane equation~\eqref{eq:mainext}
with small deflections. We assume $\partial_\alpha y \ll 1$ and $\partial_\alpha x \approx 1$. Then
~$\partial_\alpha s - 1 = \sqrt{(\partial_\alpha x)^2+(\partial_\alpha y)^2} - 1 \approx
\partial_\alpha y^2/2$ and $\hat{s}_y \approx \partial_\alpha y$, so the $y$-component of the
$R_3$ term in \eqref{eq:mainext} is cubic in deflection. By the same reasoning, the $y$-component
of the $T_0$ term is linear in deflection, and
similarly for the
$R_1$ term (multiplying $\partial_{tt} y$)---which
is zero here at steady state, but not for the
oscillating membranes considered later. The pressure
jump is linear in the bound vortex
sheet strength because the left side of \eqref{eq:pressureAlpha} $\approx \partial_t \gamma + \partial_\alpha \gamma$ with
small deflections. The bound vortex sheet
strength is linear in the deflection by the linearized version of (\ref{kinematic}), 
\begin{align}
\partial_t y(\alpha,t) &\approx \frac{1}{2\pi} \Xint-_{-1}^{1}\frac{\gamma(\alpha',t)}{x(\alpha,t)-x(\alpha',t)}\,\text{d}\alpha'-\frac{1}{2\pi}\Xint -_{0}^{\Gamma_+(t)}\frac{x(\alpha,t)-x(\Gamma',t)}{(x( \alpha,t)-x(\Gamma',t))^2+\delta(\Gamma',t)^2}\,\text{d}\Gamma',\label{kinematiclin}
\end{align}
in which the second integral consists of bound
vorticity advected from the trailing edge, so it has the same dependence on deflection as the bound vorticity.
Here, with small deflections, we have assumed
$\partial_\alpha x \approx 1$, and then the linearization is the same as in \cite{alben2008flapping}. Without viscous stresses,
horizontal membrane deformations arise only through nonlinear terms in the elastic and pressure forces associated with large deflections, so it is reasonable
to neglect them, and this is consistent with the
simulation results. Balancing the terms that are linear in deflection with the product of $R_3$ and a term that scales with deflection cubed gives $\langle y_{\mbox{defl}}\rangle\sim 1/\sqrt{R_3}$.

\subsection{Fixed-free membranes}

We now investigate membranes with the leading edge fixed and
the trailing edge free to move vertically (with $\partial_\alpha y=0$ there---an extra equation that determines $y(1,t)$, now an extra unknown). With the free end, the membrane has a wide range of unsteady dynamics with small and moderate amplitude, unlike in the fixed-fixed case, and similar in some respects to the fixed-free flag with bending rigidity \cite{alben2008flapping}. 



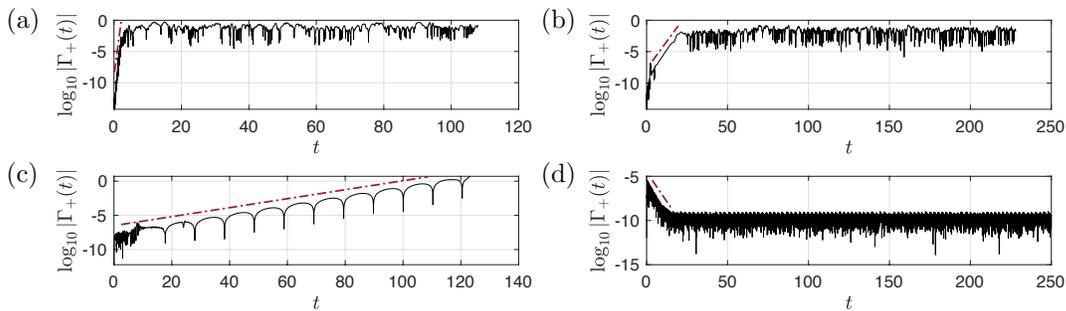
\begin{figure}[H]
	\centering
\begin{overpic}[width=.95\textwidth,tics=10]{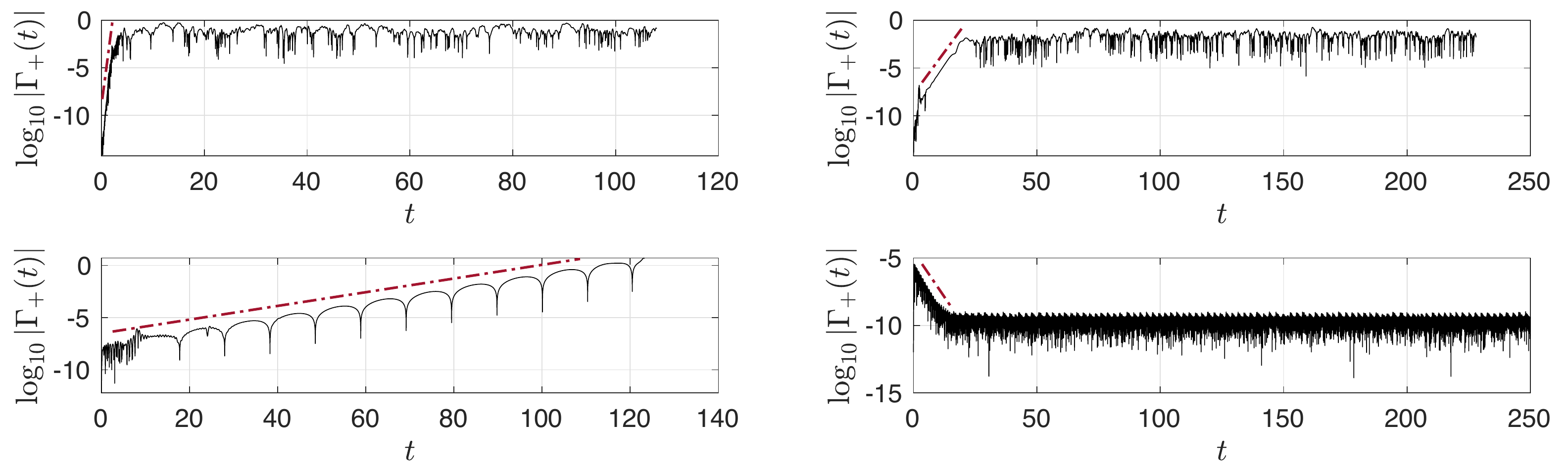}
\put(-4,28){(a)}
\put(48,28){(b)}
\put(-4,13){(c)}
\put(48,13){(d)}	
\end{overpic}
	\caption{Examples of the total wake circulation versus time for $R_3=1$ on a log scale, with (a) $(R_1,T_0)=(10^{-3},10^{-2.8})$, (b) $(R_1,T_0)=(10^{-1.4},10^{-1.4})$, (c) $(R_1,T_0)=(10^{1},10^{-0.2})$, and (d) $(R_1,T_0)=(10^{-0.8},10^{2.2})$. The slopes of the dot-dashed red lines give the growth/decay rates.}\label{fig:examplesOfGammaFree}
\end{figure}

Figure~\ref{fig:examplesOfGammaFree} shows examples of the growth of wake circulation in time after small transient perturbations, analogous
to figure \ref{fig:examplesOfGamma}. The main novelty is panel (c), an example of divergence with flutter--shown by the regularly-spaced vertical asymptotes in the logarithm of wake circulation, corresponding to an oscillatory component---exponential growth with
a complex growth rate.

\begin{figure}[H]
	\centering
	\begin{overpic}[width=.64\textwidth,tics=10]{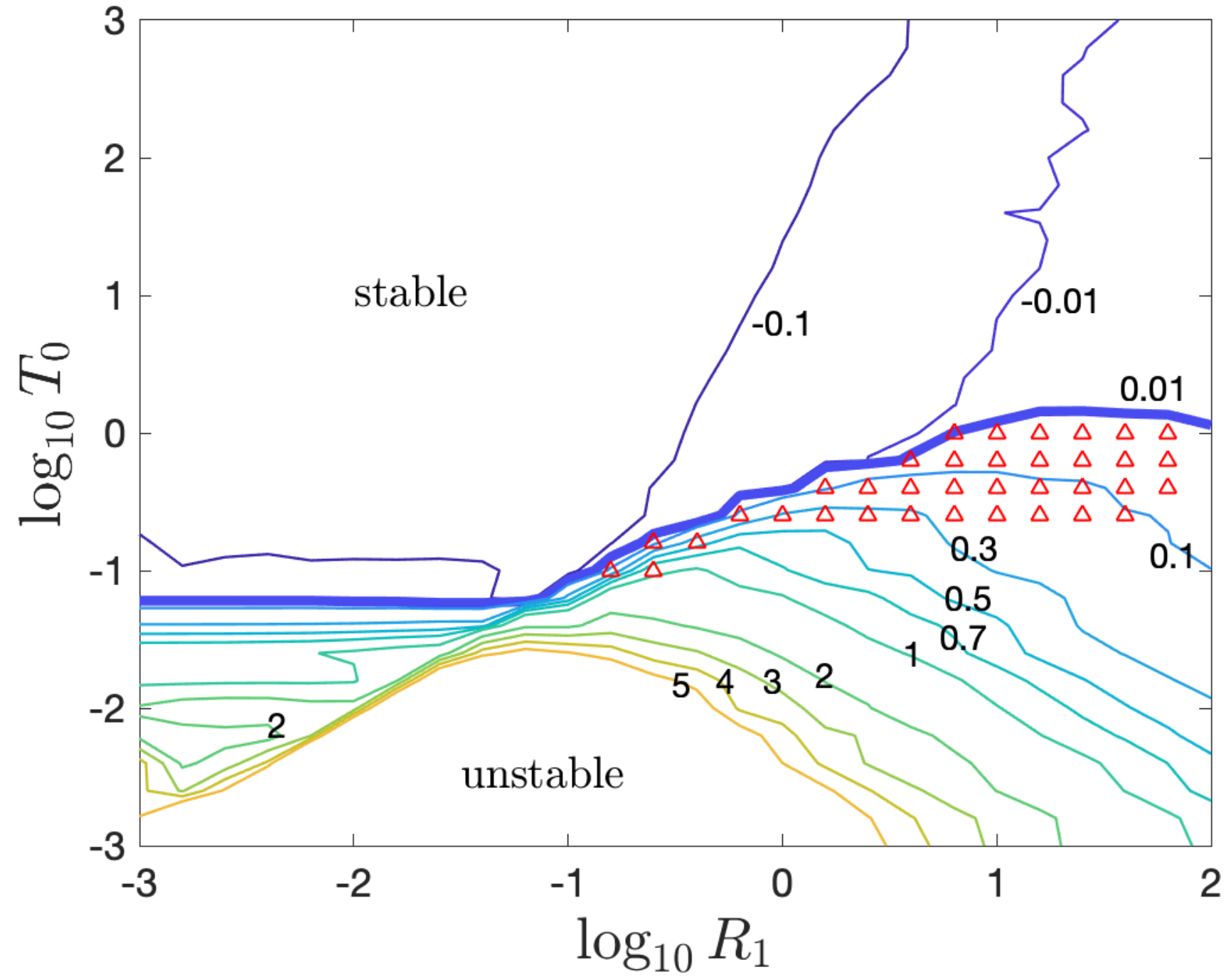}
	\end{overpic}   
	\caption{A contour plot of the exponential growth and decay rates of wake circulation after a small transient perturbation, in the fixed-free case. The thicker line separates the regions where the membranes are stable and unstable. Growth rates (well) above 5 occur at the lower left but are omitted for visual clarity.} \label{fig:stabilityFree}
\end{figure}

Figure~\ref{fig:stabilityFree} is a contour plot of the
growth/decay rates and stability boundary in $R_1$-$T_0$ space, analogous to figure~\ref{fig:stability}. Notable differences are that the stability boundary now varies with~$R_1$. The critical pretension is
close to that in figure~\ref{fig:stability} at the largest $R_1$, but
decreases as $R_1$ decreases, and eventually reaches a lower plateau
at $R_1 \ll 1$. We will show that a distinctive flapping state occurs here. The red triangles in figure~\ref{fig:stabilityFree} show cases like figure~\ref{fig:examplesOfGammaFree}(c), membranes that become unstable through flutter and divergence.


The membranes in the unstable region of figure
\ref{fig:stabilityFree} eventually reach large
amplitudes, where nonlinearities (e.g. the 
$R_3$ term) determine the eventual steady-state
motion. With fixed-free boundary conditions, oscillatory
motions are typical, unlike for the fixed-fixed case.
As for the fixed-fixed case, unrealistically large deflections occur for $R_3 \leq 1$, so we focus on~$R_3 \gtrsim 1$.



 \begin{figure}[H]
     \centering
     \includegraphics[width=.9\textwidth]{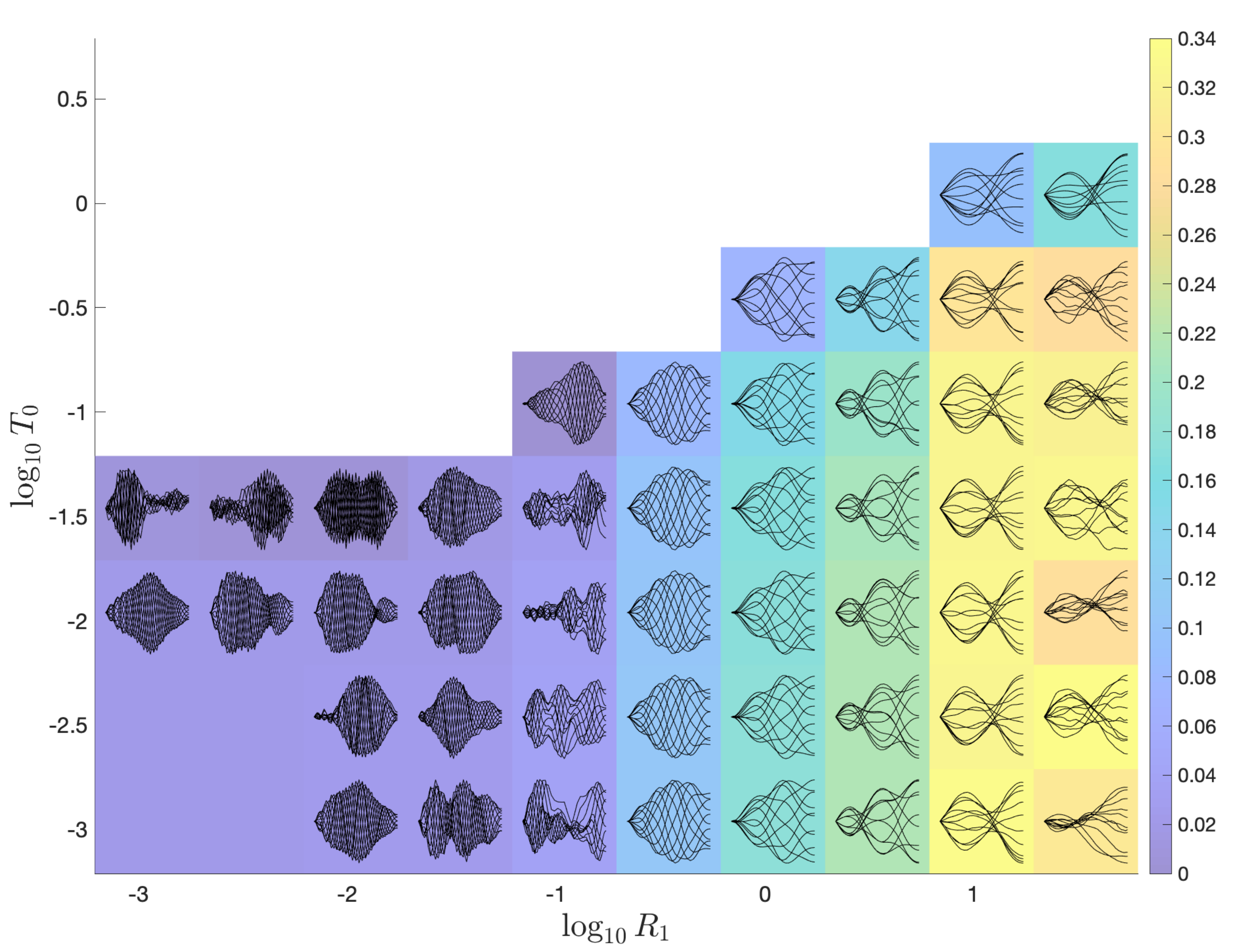}
 \caption{Snapshots of large-amplitude membrane motions in the unstable region of $R_1$-$T_0$ space, for fixed $R_3=10^{1.5}$. Colors denote  $\log_{10}$ of the average deflection defined by equation~\eqref{ydefl}.\label{fig:deflT0fixedFree}}
\end{figure}
Figure \ref{fig:deflT0fixedFree} shows typical membrane snapshots in the unstable region of $R_1$-$T_0$ space
from figure \ref{fig:stabilityFree}. 
At each ($R_1$, $T_0$) value, the set of snapshots is normalized by the maximum deflection of the snapshots and scaled to fit within a colored rectangle at the ($R_1$, $T_0$) value. Each snapshot has the corresponding $R_1$ value at its horizontal midpoint, and the $T_0$ value at its leading edge.
Here $R_3$ is fixed at $10^{1.5}$, a value giving moderately
large deflections for the steady-state motion. The colors denote the average deflection of the membrane (equation~\eqref{ydefl}).
The region in figure \ref{fig:deflT0fixedFree} can be split into two main subregions. For $R_1 < 0.1$, the membranes oscillate
with very small amplitudes and high spatial frequencies,
which do not depend much on $R_1$ because it is small enough that membrane inertia is negligible compared
to fluid pressure forces. Although the trailing edge is free, the oscillation amplitude is often very small there.
At the lower left corner, snapshots are omitted because steady-state membrane motions were not
obtained.
For $R_1 > 0.1$, the motions are more regular, with order-one amplitudes and longer wavelength components, both
of which increase with $R_1$. The motions become more
chaotic at larger~$R_1$, slightly up-down asymmetrical, with
largest deflection near the trailing edge. 
Intuitively speaking, large $R_1$ (membrane inertia) allows
the membrane to maintain its momentum for longer times against restoring fluid forces, and move farther before reversing direction. Therefore, larger amplitudes are observed.  
The membrane motion is nearly independent of $T_0$ except near the stability boundary
(at the largest $T_0$ shown). At smaller $T_0$, the pretension does not affect the dynamics because it is negligible compared to the $R_3$ (stretching) term.


\begin{figure}[H]
    \centering
    \includegraphics[width=.95\textwidth]{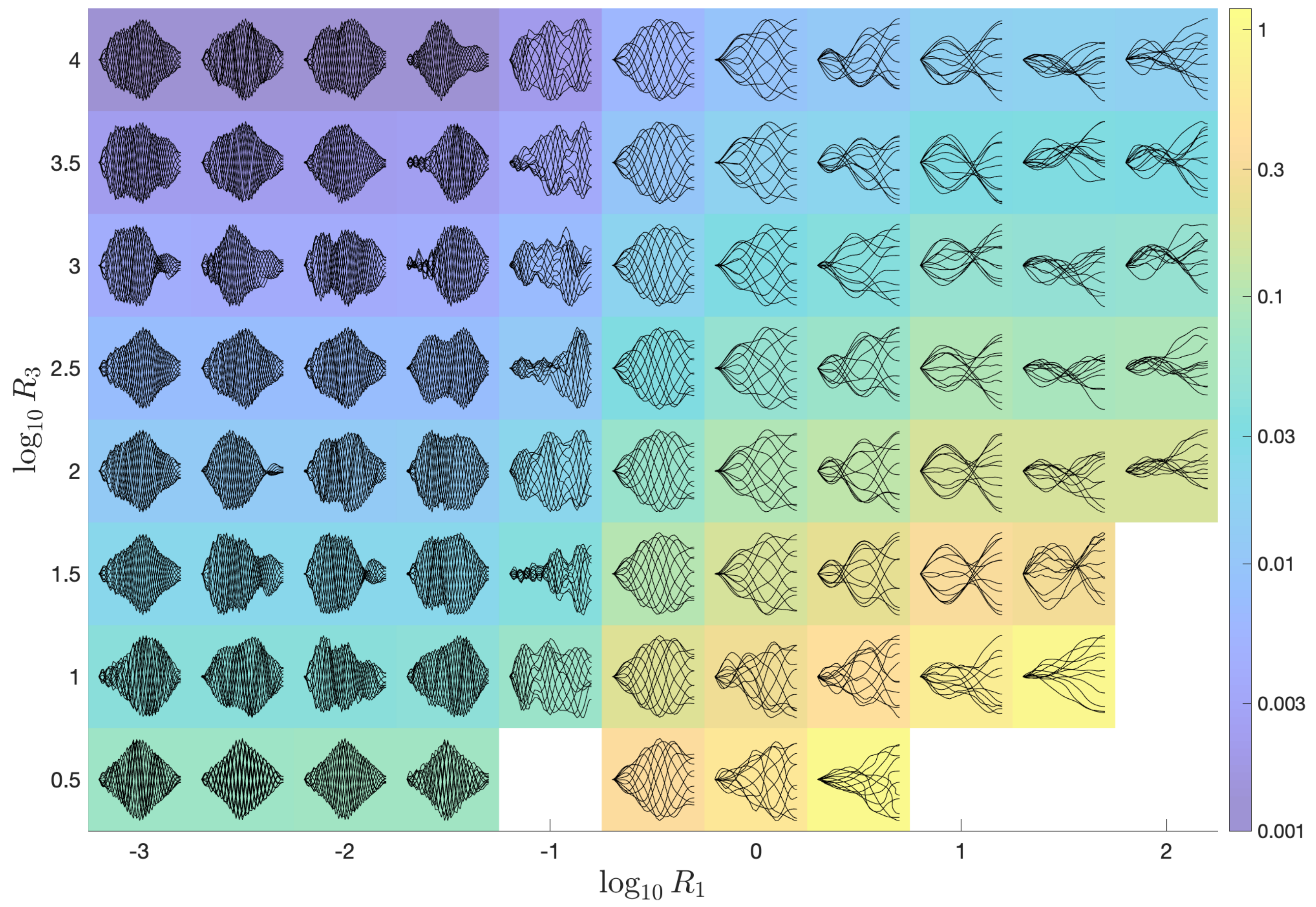}
    \caption{Snapshots of large-amplitude membrane motions in $R_1$-$R_3$ space, for fixed $T_0=10^{-2}$. Colors denote $\log_{10}$ of the average deflection defined by equation~\eqref{ydefl}.  \label{fig:deflR3fixedFree}}
\end{figure}

Next we look at the same quantities in a different two-dimensional slice through $R_1$-$T_0$-$R_3$ space. We
fix $T_0 = 10^{-2}$ and in figure \ref{fig:deflR3fixedFree} show the membrane motions across $R_1$ and $R_3$. In the lower right corner and at $(R_1,R_3)=(10^{-1},10^{0.5})$, snapshots are omitted because steady-state membrane motions were not
obtained. We see essentially the same two subregions, for large and for small $R_1$.
$R_3$ mainly affects the amplitudes of the snapshots but not their shapes, except in a few locations such as the transition zone at $R_1 = 10^{-1}$, and at $R_3 \approx 10^{0.5}$, near the smallest stretching modulus where steady-state motions occur. In these regions the motions are somewhat more irregular and more sensitive to small changes in parameters. 

\begin{figure}[H]
	\centering
	\begin{overpic}[width=.65\textwidth,tics=10]{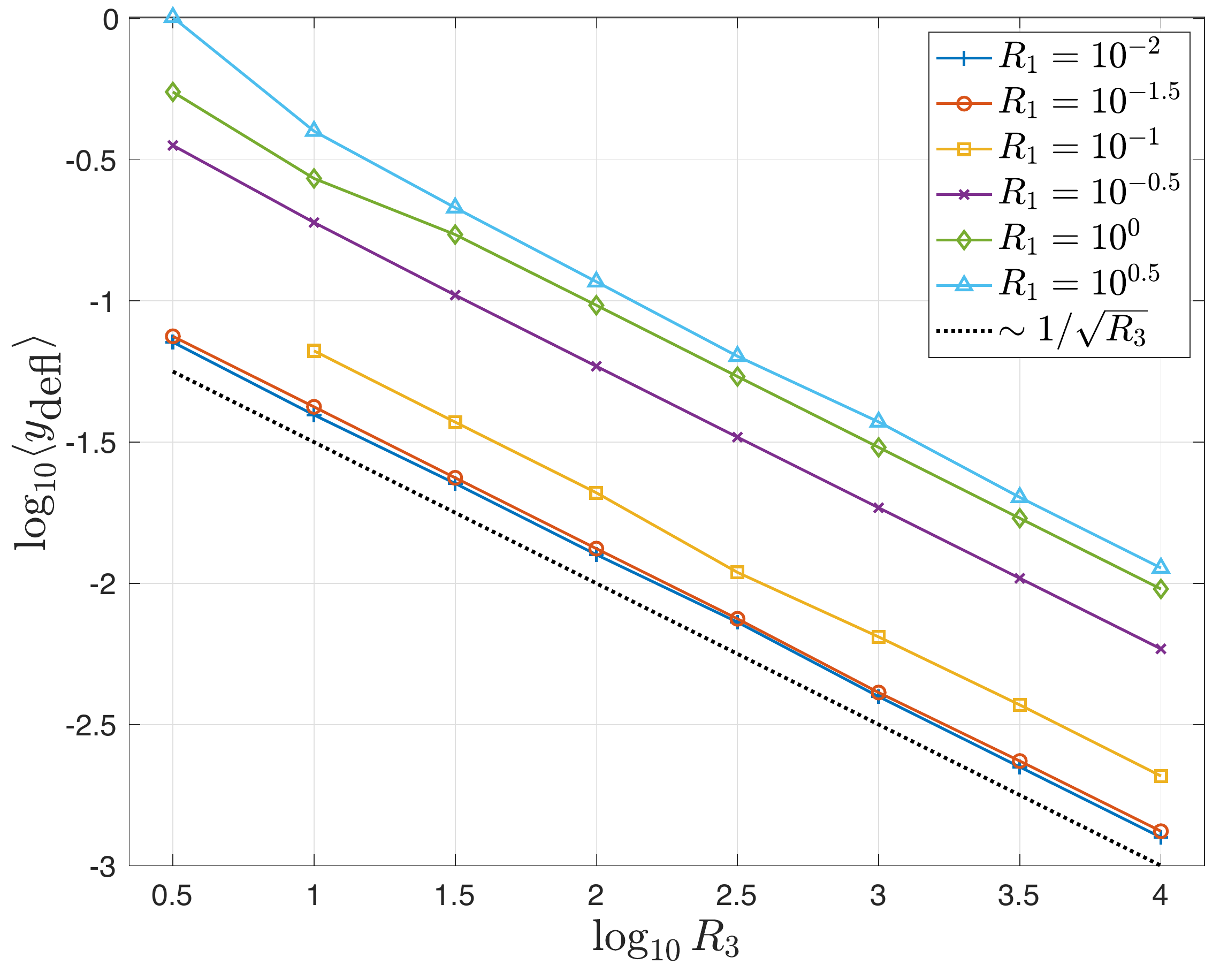}
	\end{overpic}
	\caption{Time-averaged deflections of the membranes (defined by equation~\eqref{ydefl}) versus $R_3$, for various $R_1$ and fixed $T_0=10^{-2}$. The black dashed line indicates the scaling~$1/\sqrt{R_3}$. }\label{fig:lineAvgValue}
\end{figure}

We show how the time-averaged deflection depends on $R_3$ at several fixed values
of $R_1$ in figure \ref{fig:lineAvgValue}. The plots follow the same $1/\sqrt{R_3}$
dependence at large $R_3$ as in the fixed-fixed case and for the same reason (explained in section
\ref{FixedFixed}).

Figure \ref{fig:deflR3fixedFree} has shown the
typical membrane motions at various $R_1$
and $R_3$ at fixed $T_0$ (and the phenomena are similar at other $T_0$ that yield flutter).
We now quantify the membrane shapes in terms
of the time-averaged number of ``zero crossings"---the number of times the membrane crosses $y = 0$. This is one way to measure the ``waviness" of a shape which is not sinusoidal (so the wavelength is not
well defined) \cite{alben2015flag,alben2008flapping}. 

\begin{figure}[H]
\centering
\begin{overpic}[width=.7\textwidth,tics=10]{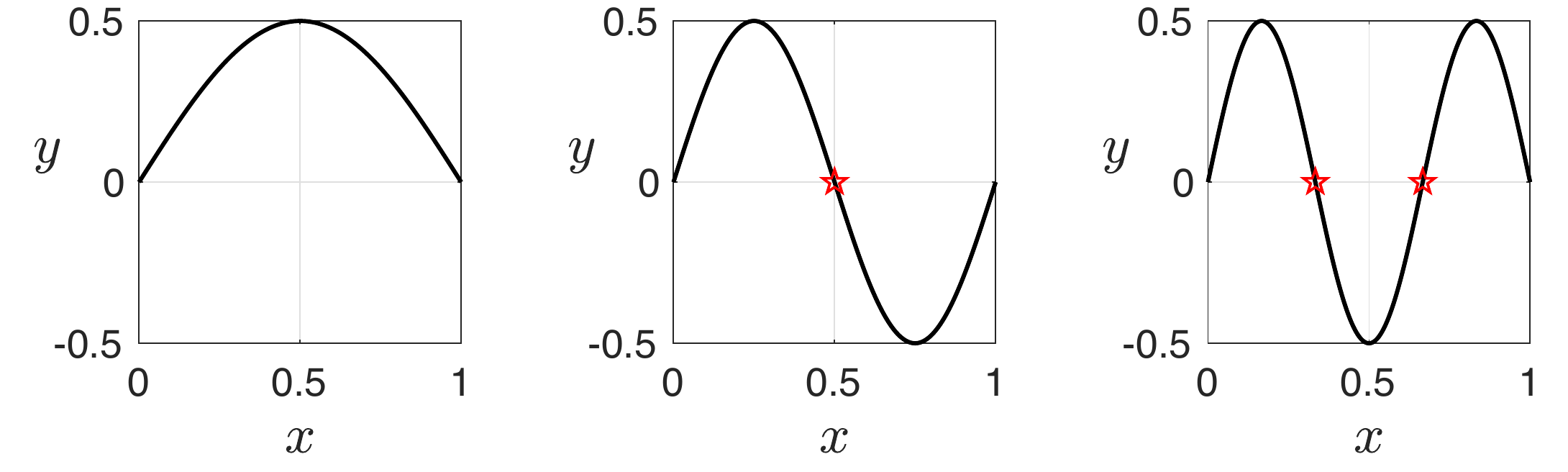}
\end{overpic}
\caption{Examples of where zero crossings (red stars) are counted for model membranes (black). Note that the endpoints are not included.}\label{fig:schematicZeroCrossings}
\end{figure}

Figure \ref{fig:schematicZeroCrossings} shows
examples of shapes with zero, one, and two zero-crossings, respectively. The number of zero
crossings is usually, but not always, one less than the number of local extrema for the membrane motions
we have observed.

\begin{figure}[H]
	\centering
	\includegraphics[width=.95\textwidth]{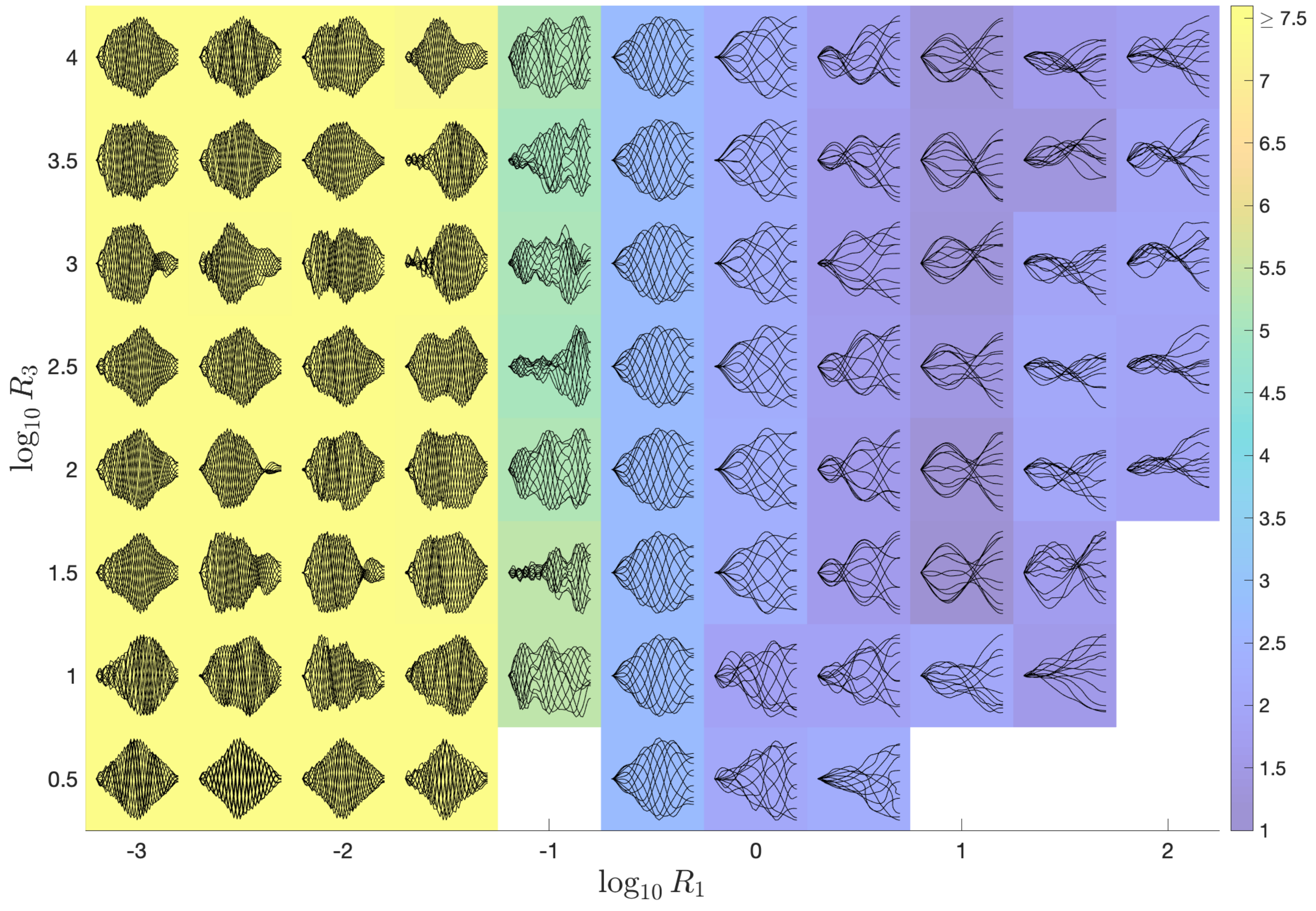}
	\caption{Colors denote the time-averaged number of zero-crossings for membrane flutter in the $R_1$-$R_3$ parameter space, for fixed $T_0=10^{-2}$. In the lower right corner and at $(R_1,R_3)=(10^{-1},10^{0.5})$, snapshots are omitted because steady-state membrane motions were not
obtained.}\label{fig:snapshotsFixedFree}
\end{figure}

Figure~\ref{fig:snapshotsFixedFree} shows that the average number of zero crossings is large ($\geq 7$) in the region $R_1 < 10^{-1}$ (where fluid inertia dominates), and much smaller for
$R_1 > 10^{-1}$, decreasing from~3 to about 1 as $R_1$ increases to $10^2$. 

\begin{figure}[H]
	\centering
	\includegraphics[width=.95\textwidth]{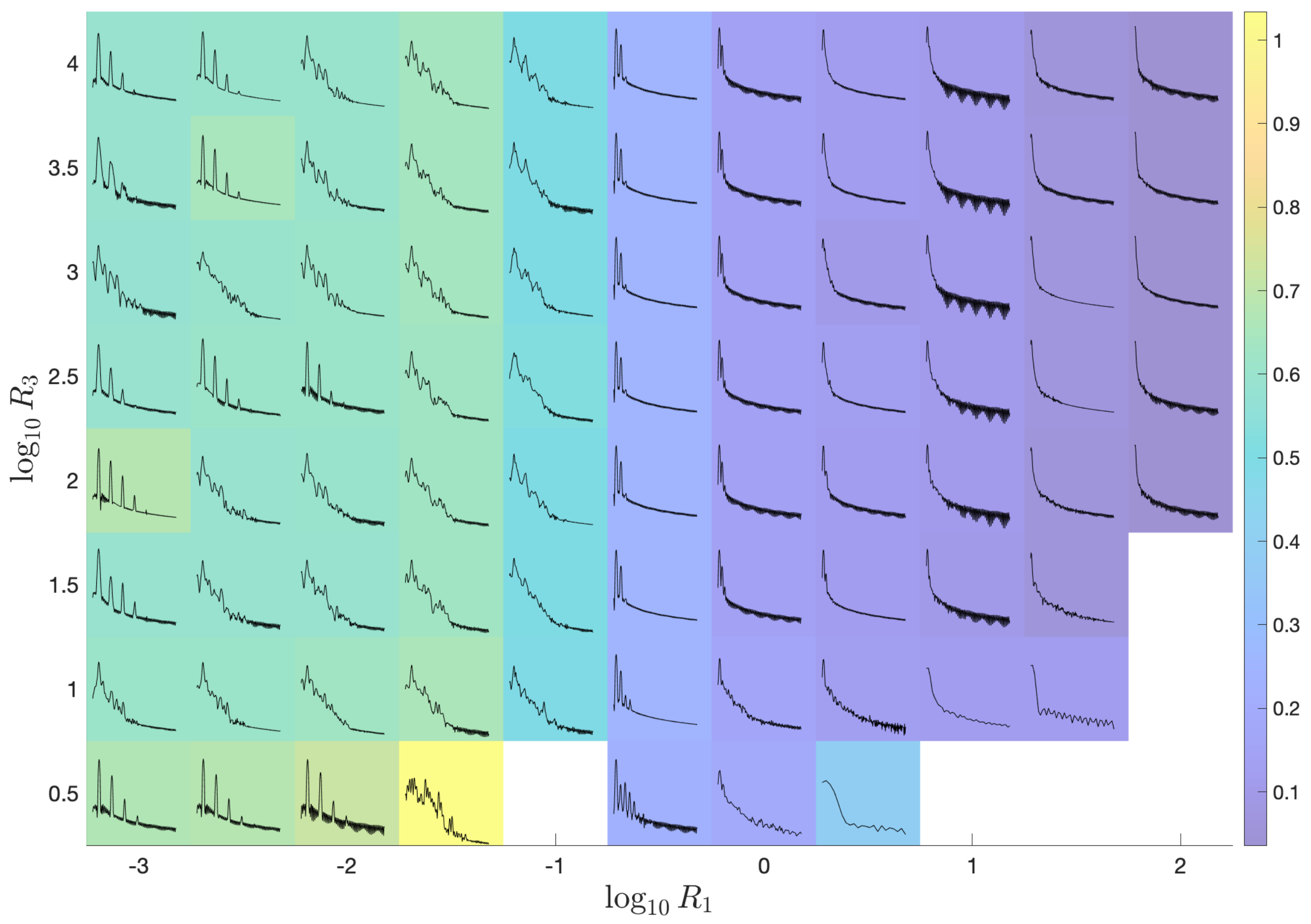}
	\caption{Colors denote the mean frequencies of large-amplitude motions in the fixed-free case for various $R_1$ and
	$R_3$, all with $T_0=10^{-2}$. The corresponding power spectra for each membrane are plotted in black. In the lower right corner and at $(R_1,R_3)=(10^{-1},10^{0.5})$, power spectra are omitted because steady-state membrane motions were not
obtained.}\label{fig:periodogram}
\end{figure}

The temporal dynamics corresponding to these motions are quantified by computing the power spectra of
time series of the total wake circulation, $\Gamma_+(t)$. Figure \ref{fig:periodogram}
shows these spectra, computed using Welch's method
\cite{welch1967use}, in $R_1$-$R_3$ space, with $T_0=10^{-2}$. The colors denote the
mean frequencies---i.e., the first moments of the
power spectra, normalized by total power. In the region $R_1 < 10^{-1}$, the mean frequencies are mostly 0.6--0.7, about 2--10 times larger than those
at $R_1 > 10^{-1}$. Thus there is a strong correlation between number of zero-crossings (or flutter mode) and oscillation frequency, as has been
seen previously in flag flutter problems \cite{shelley2005heavy,ESS2007,alben2008flapping,alben2015flag}. The two are in linear proportion for the
modes of the linear wave equation for a membrane in a vacuum \cite{graff1975wave,farlow1993partial}.
The power spectra for $R_1 < 10^{-1}$
are mostly dominated by a few peaks, harmonics
of a dominant flapping frequency. The peaks may be
very sharp (as for $R_1 = 10^{-2.5}$ and $R_3 \geq 10^{3.5}$),
indicating oscillation at a particular frequency,
or more spread out (as for $R_1 = 10^{-1.5}$) indicating a less periodic motion, also
indicated by the greater irregularity of the snapshots in figure~\ref{fig:snapshotsFixedFree}
at $R_1 = 10^{-1.5}$. As $R_1$ increases from
$10^{-1}$ to $10^{-0.5}$, the spectral peaks become sharper, indicating a transition to a regular periodic motion. With further increase of $R_1$, the spectra gradually broaden, consistent with the less regular motions at the largest $R_1$.

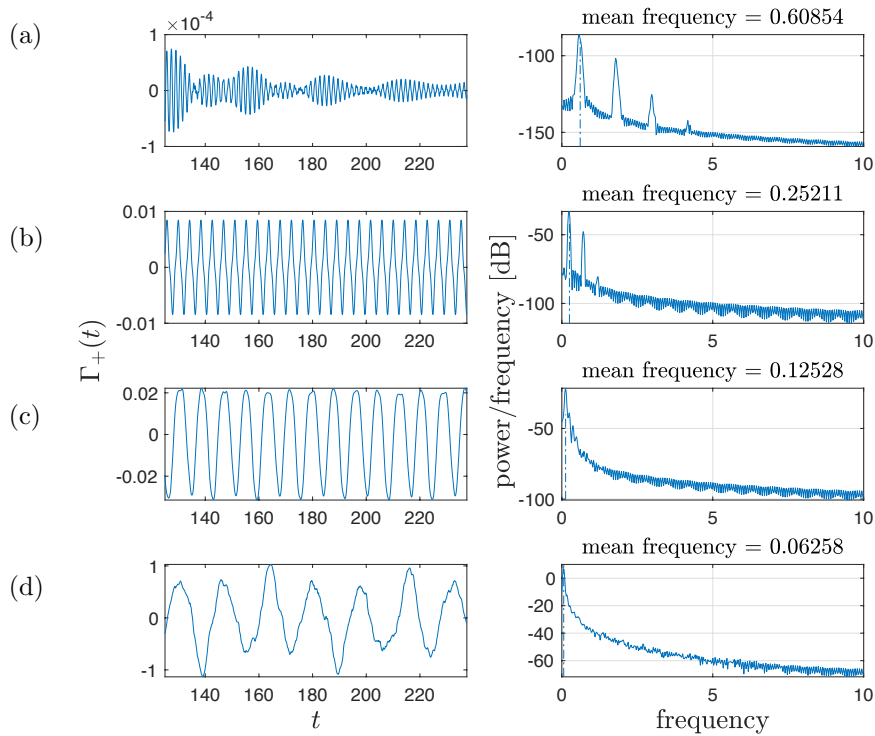
\begin{figure}[H]
\centering
\begin{overpic}[width=.75\textwidth,tics=10]{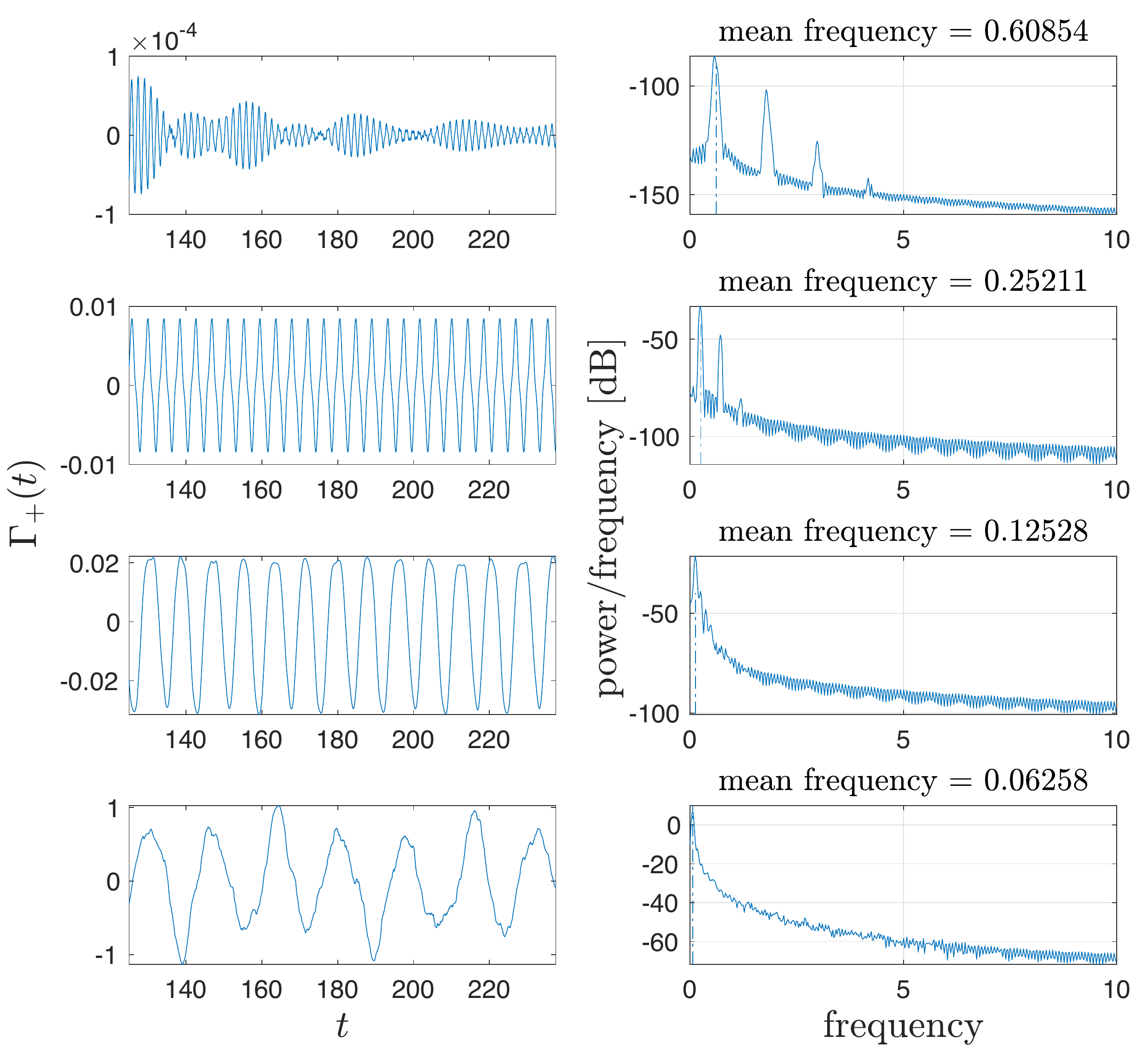}
\put(-8,86){(a)}
\put(-8,61){(b)}
\put(-8,39){(c)}
\put(-8,18){(d)}
\end{overpic}
\caption{Portion of time series of total wake circulation $\Gamma_+(t)$ (left) and corresponding power spectra (right) for membranes with increasing
mass density $R_1$ from top to bottom:
(a) $(R_1,R_3)=(10^{-2.5},10^{3.5})$, (b) $(R_1,R_3)=(10^{-0.5},10^{2})$, (c) $(R_1,R_3)=(10^{0.5},10^{3.5})$, and (d) $(R_1,R_3)=(10^{1.5},10^{1.5})$. In all cases $T_0=10^{-2}$.  }\label{fig:period}
\end{figure}

Figure~\ref{fig:period} shows examples of the time series of $\Gamma_+(t)$ that correspond to the different types of power spectra. Panel (a) is typical of those with $R_1 < 10^{-1}$, and shows a high-frequency oscillation with a low-frequency envelope. This corresponds to a spectrum dominated by a small band of frequencies spread near the lowest peak and its higher harmonics. Panel (b) shows an example just above the transition to larger $R_1$ dynamics, with a more periodic (though not sinusoidal) motion, corresponding to sharper peaks in the power spectrum.
Panel (c) shows a less periodic response at larger $R_1$, still dominated by a single frequency but with clear variations from one cycle to the next. Panel (d) shows the trend toward aperiodicity continuing at larger $R_1$, with little evidence of discrete peaks in the power spectrum. Nonetheless, the time series (left) shows peaks with a somewhat regular spacing.

\begin{figure}[H]
\centering
\begin{overpic}[width=.7\textwidth,tics=10]{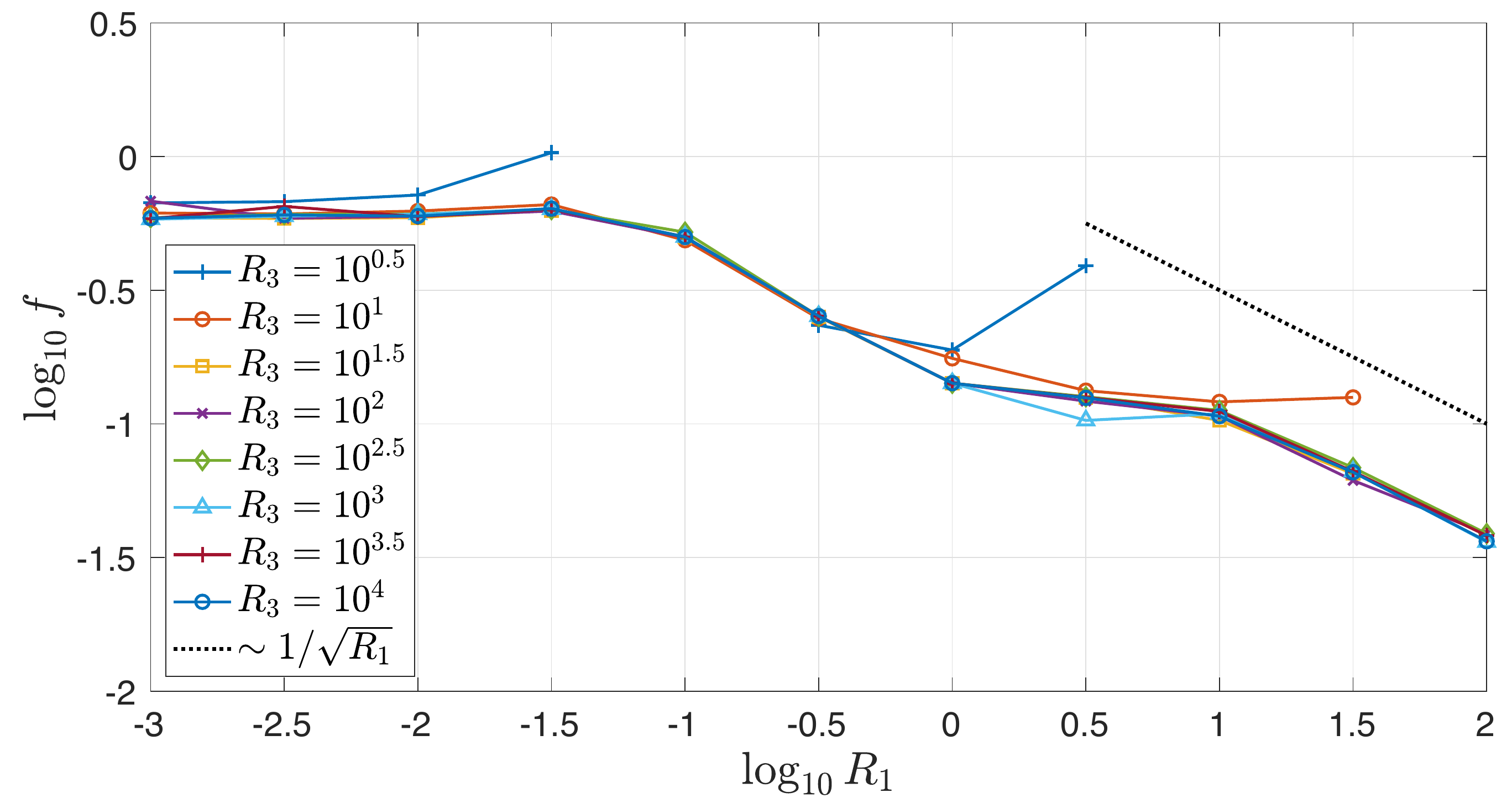}
\end{overpic}
\caption{Plots of the mean frequency $\log_{10} f$ versus mass density $\log_{10} R_1$ with various $R_3$ and fixed $T_0=10^{-2}$. The black dashed line shows $f = 1/\sqrt{R_1}$.}\label{fig:lineFrequency}
\end{figure}

Figure~\ref{fig:lineFrequency} shows more quantitatively how the mean frequency varies with parameters. There is very little dependence on $R_3$ except near the smallest $R_3$ where stable motions can be computed, $\approx 10^{0.5}$. As already noted, there is little variation in the
$R_1 < 10^{-1}$ regime, then a sharp drop from $R_1 = 10^{-1}$ to $10^{0}$ 
followed by a small plateau for
$10^0 \leq R_1 \leq 10^{1}$, and another drop within
$10^1 \leq R_1 \leq 10^{2}$. The trend at the
largest $R_1$ is well approximated by 
$f \sim 1/\sqrt{R_1}$ (admittedly over a short range of $R_1$), except at the two smallest $R_3$ values. This scaling arises when one approximates the normal component of the membrane equation (\ref{eq:mainext}) by its~$y$ component, and chooses a characteristic time scale $t_0$ so that $R_1 \partial_{tt} y$ balances other terms that depend on~$y$ but not its time-derivatives (i.e. the $R_3$ and $T_0$ terms and some of the fluid pressure terms). 
At large $R_1$, $R_1 \partial_{tt} y$ is comparable to the other terms
when $R_1/t_0^2 \sim 1$ or $t_0 \sim \sqrt{R_1}$,
giving a typical frequency $f_0 \sim 1/\sqrt{R_1}$.

\begin{figure}[H]
	\centering
	\includegraphics[width=.95\textwidth]{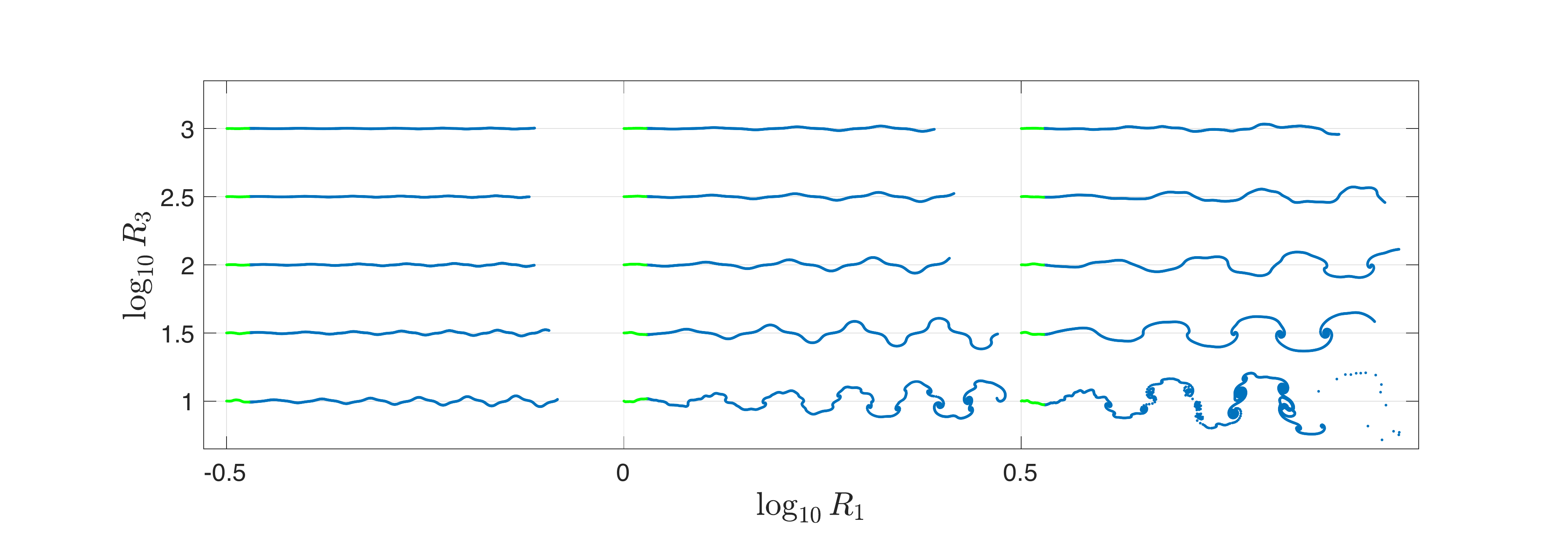}
	\caption{Snapshots of the membrane motion (in green) and the vortex wake (in blue) in a portion
	of the $R_1$-$R_3$ parameter space for fixed $T_0=10^{-2}$. In each case the ($R_1$, $R_3$) values are marked at the left endpoint of the membrane. }\label{fig:vortex2DFixedFree}
\end{figure}

We have mainly focused on the membrane dynamics, but we conclude this section by briefly considering the vortex sheet wake dynamics. Figure 
\ref{fig:vortex2DFixedFree} shows snapshots of
vortex wakes in a small portion of $R_1$-$R_3$
space where the membranes' motions transition from small to large amplitudes. At smaller amplitudes (top and left), the wakes are mostly flat (despite the
complexity of the corresponding membrane snapshots, shown in figure \ref{fig:deflR3fixedFree}), and
have periodic undulations, as the membrane motions
are approximately periodic at these parameters.
At the smallest $R_3$ and largest $R_1$ (bottom right), the wakes become more complex, corresponding
to less regular, larger-amplitude membrane motions.






\subsection{Free-free membranes}

We have seen that changing the trailing edge boundary condition from fixed to free dramatically changes
the membrane dynamics, from static deflections
with a single maximum to a wide range of oscillatory modes that have some commonalities with 
flapping plates and flags 
\cite{shelley2011flapping}. Therefore it is natural
to consider the effect of making both ends free, and
determine if the membrane dynamics undergo further
dramatic changes. Computationally, the method is the same as before, but the system of unknowns now includes the values of $y$ at both endpoints, corresponding to the two equations $\partial_\alpha y(-1,t)=\partial_\alpha y(1,t)=0$. 

\begin{figure}[H]
	\centering
	\begin{overpic}[width=.64\textwidth,tics=10]{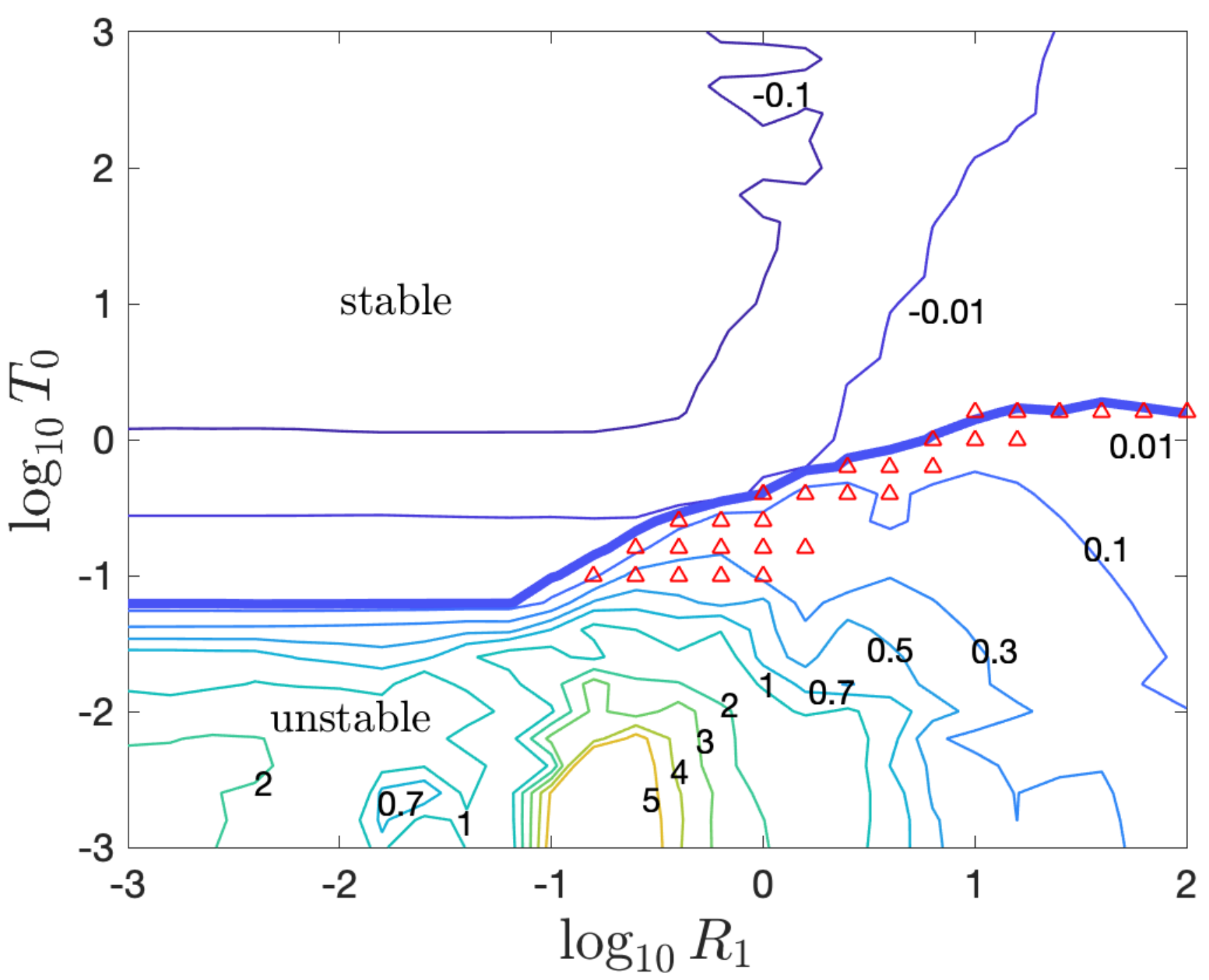}
	\end{overpic}
	\caption{A contour plot of the exponential growth and decay rates of wake circulation after a small transient perturbation, in the free-free case. The thicker line separates the stable membranes from the unstable ones.} \label{fig:stabilityFreeFree}
\end{figure}
The contour plot of growth/decay rates of the initial perturbation is shown in figure
\ref{fig:stabilityFreeFree}. It is similar
to that in the fixed-free case (figure \ref{fig:stabilityFree}), particularly in the location of the stability boundary and the contours in the stable region. In the unstable region, the growth rates are significantly smaller and there are slight differences in where divergence with flutter occurs (red triangles).



\begin{figure}[H]
	\centering
	\begin{overpic}[width=.95\textwidth,tics=10]{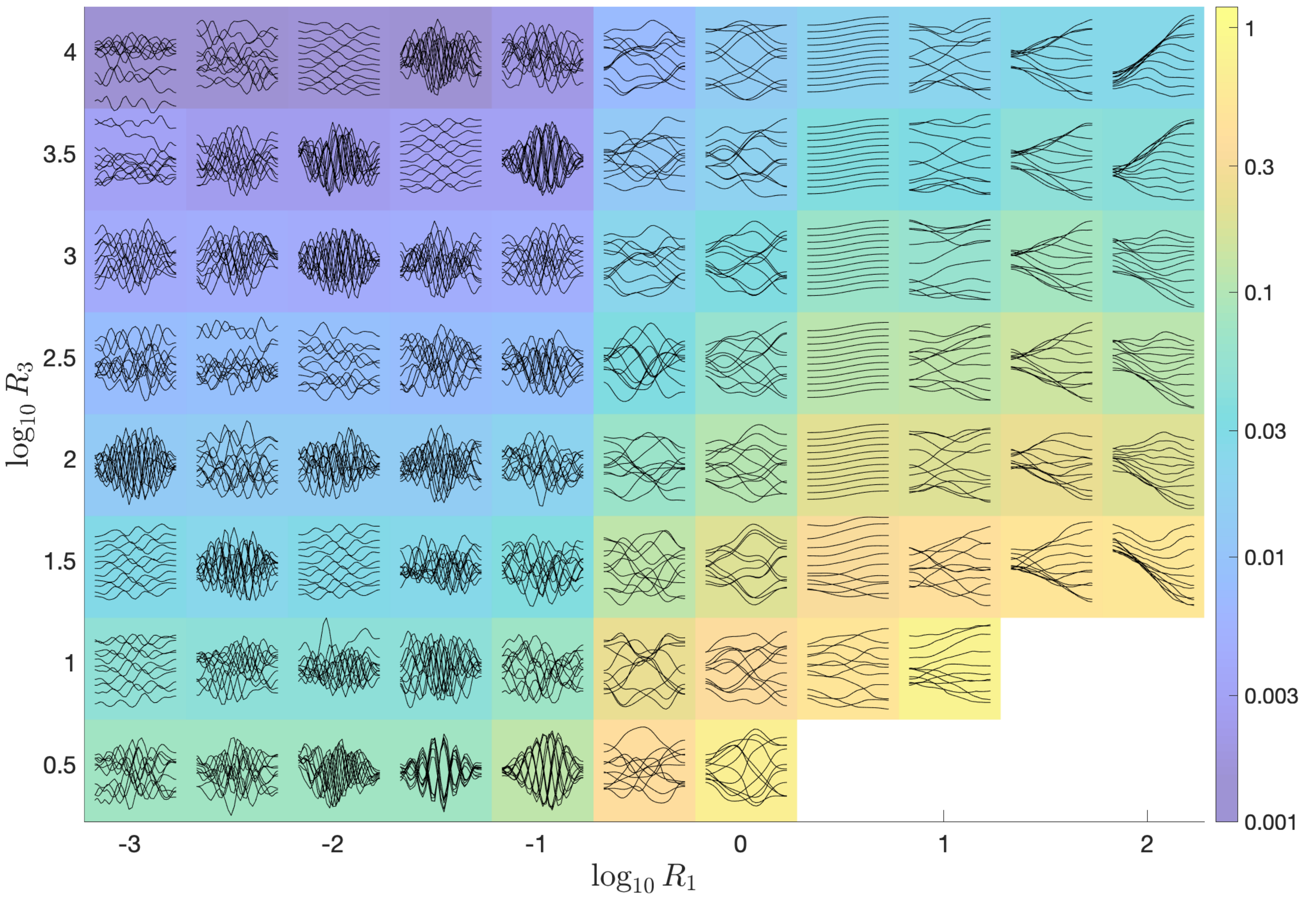}
	\end{overpic}
	\caption{Membrane snapshots in the free-free case superposed on colors labeling the time-averaged deflection for each ($R_1,R_3$) pair with $T_0=10^{-2}$ defined by equation~\eqref{ydefl}. }\label{fig:deflectionFreeFree}
\end{figure}

We now consider the large-amplitude membrane motions
in $R_1$-$R_3$ space with $T_0$ fixed at $10^{-2}$, the free-free analog of figure \ref{fig:deflR3fixedFree}. Figure \ref{fig:deflectionFreeFree} shows
the motions superposed on a color field that
labels the time-averaged membrane deflections \eqref{ydefl}.
Like the fixed-free case (figure~\ref{fig:deflR3fixedFree}), the snapshots in figure~\ref{fig:deflectionFreeFree}
can be divided into two subregions.
For $R_1 \leq 0.1$ the shapes generally
have smaller deflections,
and many more spatial oscillations.
For $R_1 > 0.1$, the shapes have larger
deflections that grow
with $R_1$. The membrane is now free to translate in the $y$-direction, which leads to additional complexities in the motions. For $R_1 \leq 0.1$, some of the membranes translate farther upwards or downwards (e.g.\
$R_1 = 10^{-3}$ and $10^{-2}$ at
$R_3 = 10^{1.5}$) than others---which are more similar
to those in figure~\ref{fig:deflR3fixedFree}
at $R_1 < 0.1$. In the region $R_1 > 0.1$ the membranes mostly oscillate within a fixed
vertical region, except at $R_1 = 10^{0.5}$,
where the membranes mostly translate
steadily in $y$, and with a steady shape. Here the
membranes are somewhat straighter than
in the fixed-free case. The membrane deflections generally decrease with $R_3$, as previously.

\begin{figure}[H]
	\centering
	\begin{overpic}[width=.95\textwidth,tics=10]{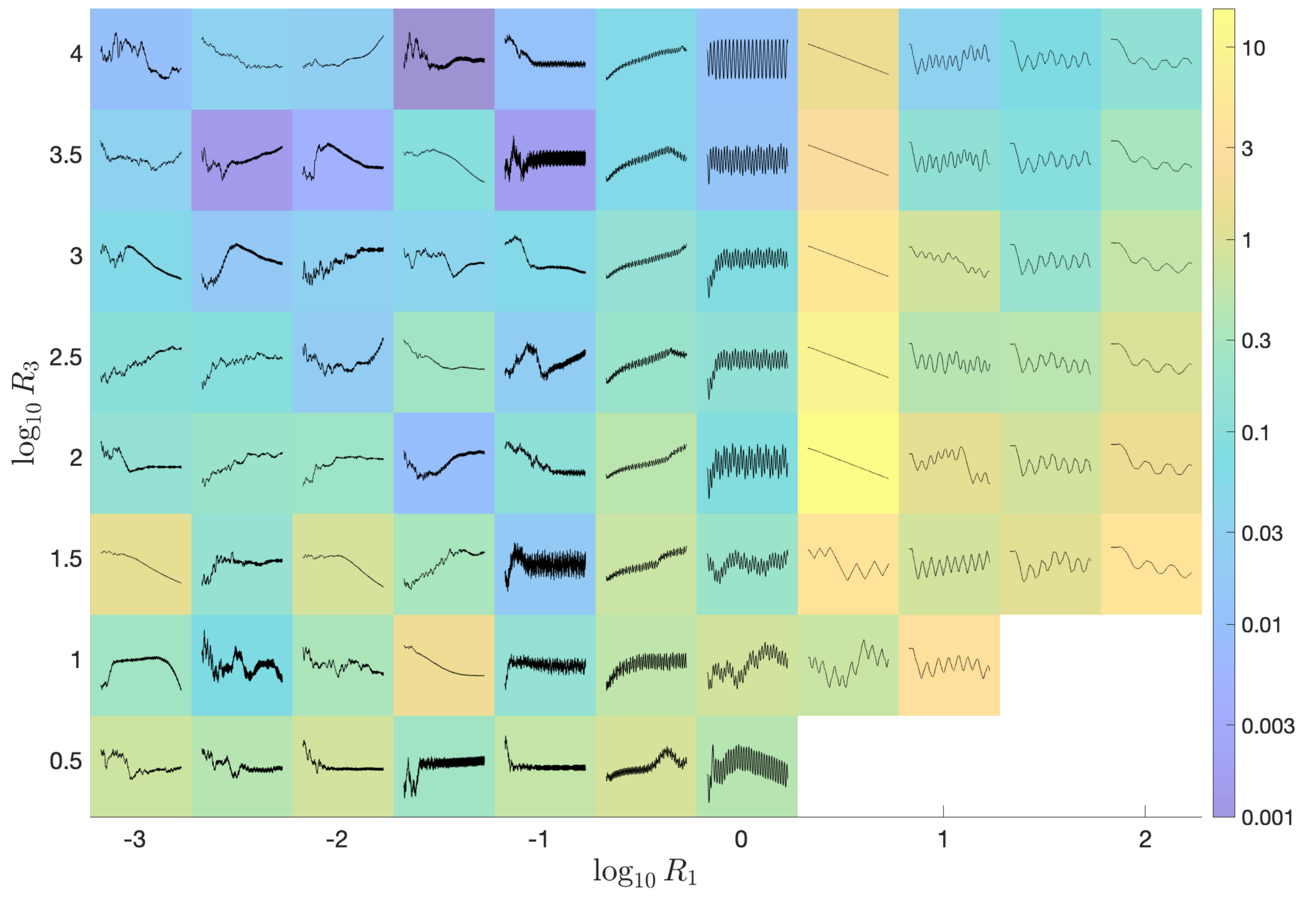}
	\end{overpic}
	\caption{Time series of the $y$ coordinates of the free-free membranes' midpoints, superposed on
	colors giving the maximum values of the time series for each ($R_1,R_3$) pair with $T_0=10^{-2}$. In the lower right corner, data are omitted because long-time trajectories were not obtained.}\label{fig:avgValueFreeFree}
\end{figure}

To show the net translational
motions of the membranes, we plot
in figure \ref{fig:avgValueFreeFree} the $y$ coordinates of the membranes' midpoints
over time. The color denotes the net $y$ displacement up to $t = 250$. For $R_1 \leq 0.1$ the vertical translations
are often chaotic, meandering upwards and downwards at irregular intervals.
In many cases however, the translation is
mostly in one direction. At $R_1 = 0.1$,
there are several cases of periodic oscillation in $y$ without much net
translation (particularly at the smallest
and largest $R_3$). Increasing
$R_1$ to $10^{-0.5}$, we have many cases of oscillation superposed on a steady
translation. At $R_1 = 1$,
essentially periodic trajectories occur (at this
$R_1$, periodic motions were seen in the fixed-free case, figure
\ref{fig:deflR3fixedFree}). At $R_1 = 10^{0.5}$, the membranes translate steadily (with occasional changes in direction, at $R_3 = 10^1$ and $10^{1.5}$). As $R_1$ increases further, the motions are mostly oscillatory with less regularity and decreasing frequencies.
The net membrane translations generally decrease with~$R_3$, presumably because
the membranes are flatter, so they have a more tangential motion
with respect to the oncoming flow if their
vertical translations are smaller, and
tangential motions are not resisted in
this inviscid model.



We can use the power spectra of the total
wake circulation to again characterize the membranes' temporal dynamics in
the same $R_1$-$R_3$ parameter space. We find properties that are similar to the fixed-free case: high frequency, chaotic motions for $R_1 < 0.1$, a sharp transition to much lower-frequency periodic motions near $R_1 = 0.1$, and
then further decreases in mean frequency but increasingly aperiodic motions at larger~$R_1$. The diverse types of translational motions do not lead to large qualitative changes in the power spectra, except for the steadily translating motions, which have only a zero-frequency component in the wake
circulation (the total wake circulation decays to zero in these cases). We present power spectrum data
for the free-free case in appendix 
\ref{freefreefreq}.

\begin{figure}[H]
\centering
\begin{overpic}[width=.95\textwidth,tics=10]{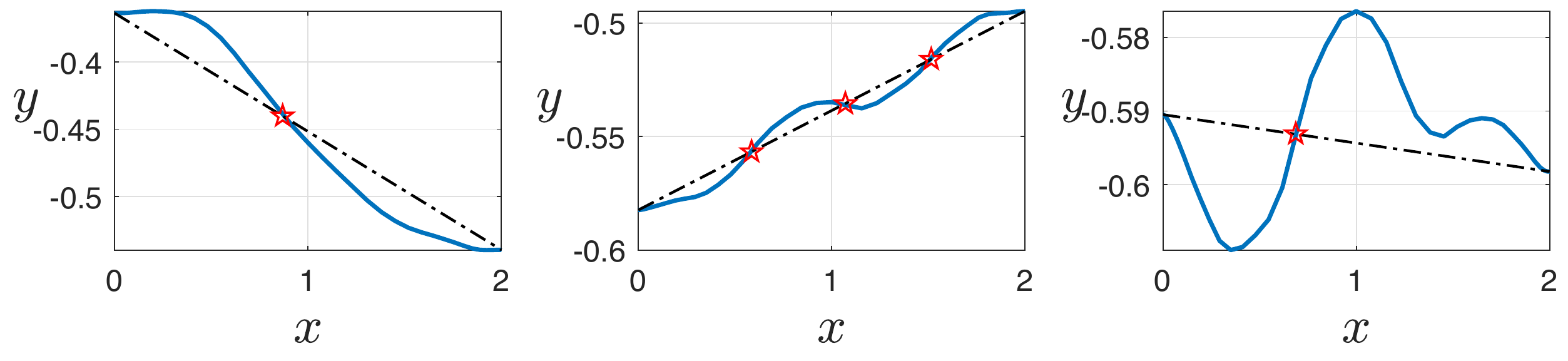}
\end{overpic}
\caption{Schematic diagram that explains the term \textit{zero-crossings} for a membrane with both endpoints free. The dot-dashed black line is the linear line that connects the two endpoints of the membrane at each time step, the blue solid line resembles the membrane at an instant of time, and the red star denotes the zero-crossing, which is the intersection point between the linear line and the membrane profile.}\label{fig:schematicZeroCrossingsFreeFree}
\end{figure}

When both membrane ends are free, the ``waviness" of the membrane is more difficult to define. Our definition is the number of crossings that a membrane makes with the line connecting its two endpoints, averaged over time. A definition based on crossings of a horizontal line would ignore the fact that many of the free-free membranes have small undulations about a line with nonzero net slope. The combination of a steady background flow with a nearly steady vertical translation makes a line with nonzero slope the state of pure tangential motion relative to the fluid, and thus the basic state of minimal resistance to the fluid.
Figure~\ref{fig:schematicZeroCrossingsFreeFree} illustrates the zero crossings using this definition for several membrane examples. Again we omit the two endpoints from the set of zero crossings.

\begin{figure}[H]
	\centering
	\includegraphics[width=.95\textwidth]{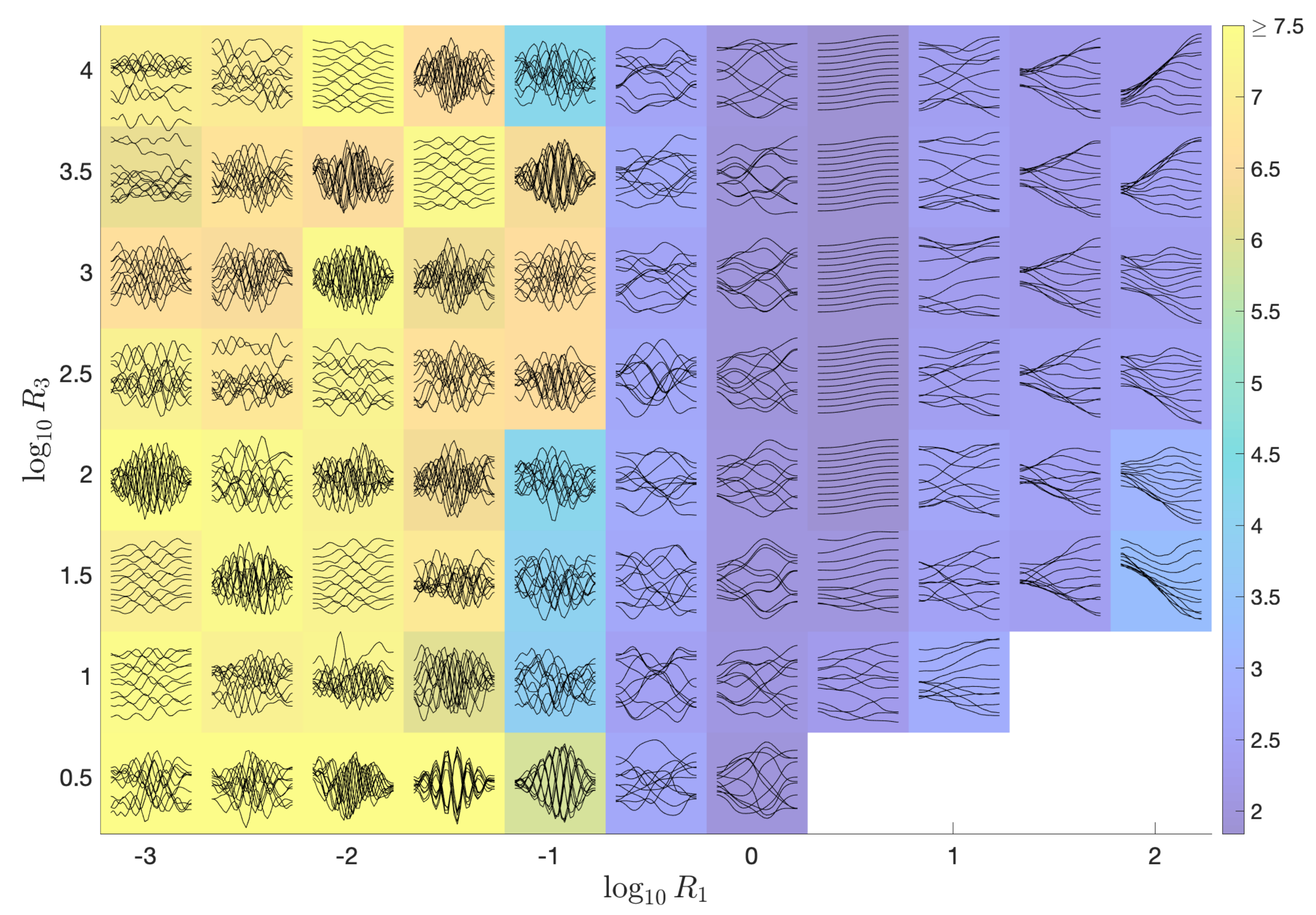}
	\caption{Snapshots of the membrane motion at the large-amplitude regime, plotted on top of colors that correspond to the number of zero-crossings in the $R_1$-$R_3$ parameter space for fixed $T_0=10^{-2}$. The data in the right bottom corner are obtained for a shorter time and so, we neglect the computational results for those values of $R_1$ and $R_3$.}\label{fig:snapshotsFreeFree}
\end{figure}

Figure~\ref{fig:snapshotsFreeFree} shows the average number
 of zero crossings in the $R_1$-$R_3$ space already discussed. Like the power spectra, this measure of membrane motion
filters out some of the differences
in translational motion. There is a clearer division of the membrane shapes into the two subregions, the one with $R_1 < 0.1$ having 6--7 zero crossings typically, and the other, with $R_1 > 0.1$ having 2--3 zero crossings typically.
There is actually a slight increase in
the number of zero crossings with $R_1$
at the largest $R_1$, as the membranes' shapes become almost straight, but with small undulations about the lines connecting the ends. Some of the
membranes at the transitional 
$R_1 = 0.1$---those with $R_3 = 10^{1},10^{1.5},10^{2}$, and $10^{4}$---have fewer average zero-crossings than the others. These membranes assume shapes akin to that in the rightmost plot of figure~\ref{fig:schematicZeroCrossingsFreeFree}. The membranes have undulations that do not pass through the lines connecting the endpoints.


\begin{figure}[H]
	\centering
	\includegraphics[width=.95\textwidth]{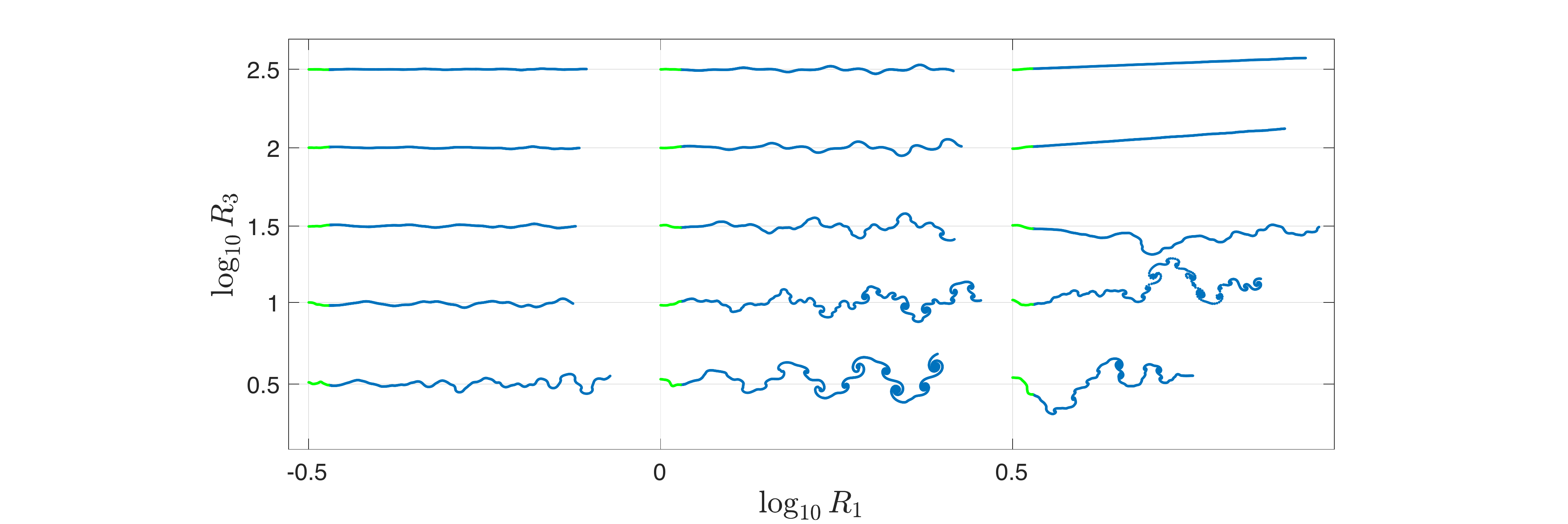} 
	\caption{Snapshots of the membrane motion (in green) and the vortex wake (in blue) at the large-amplitude regime in the $R_1$-$R_3$ parameter space for fixed $T_0=10^{-2}$.}\label{fig:vortex2DFreeFree}
\end{figure}

\begin{figure}[H]
\centering
	\begin{overpic}[width=.42\textwidth,tics=10]{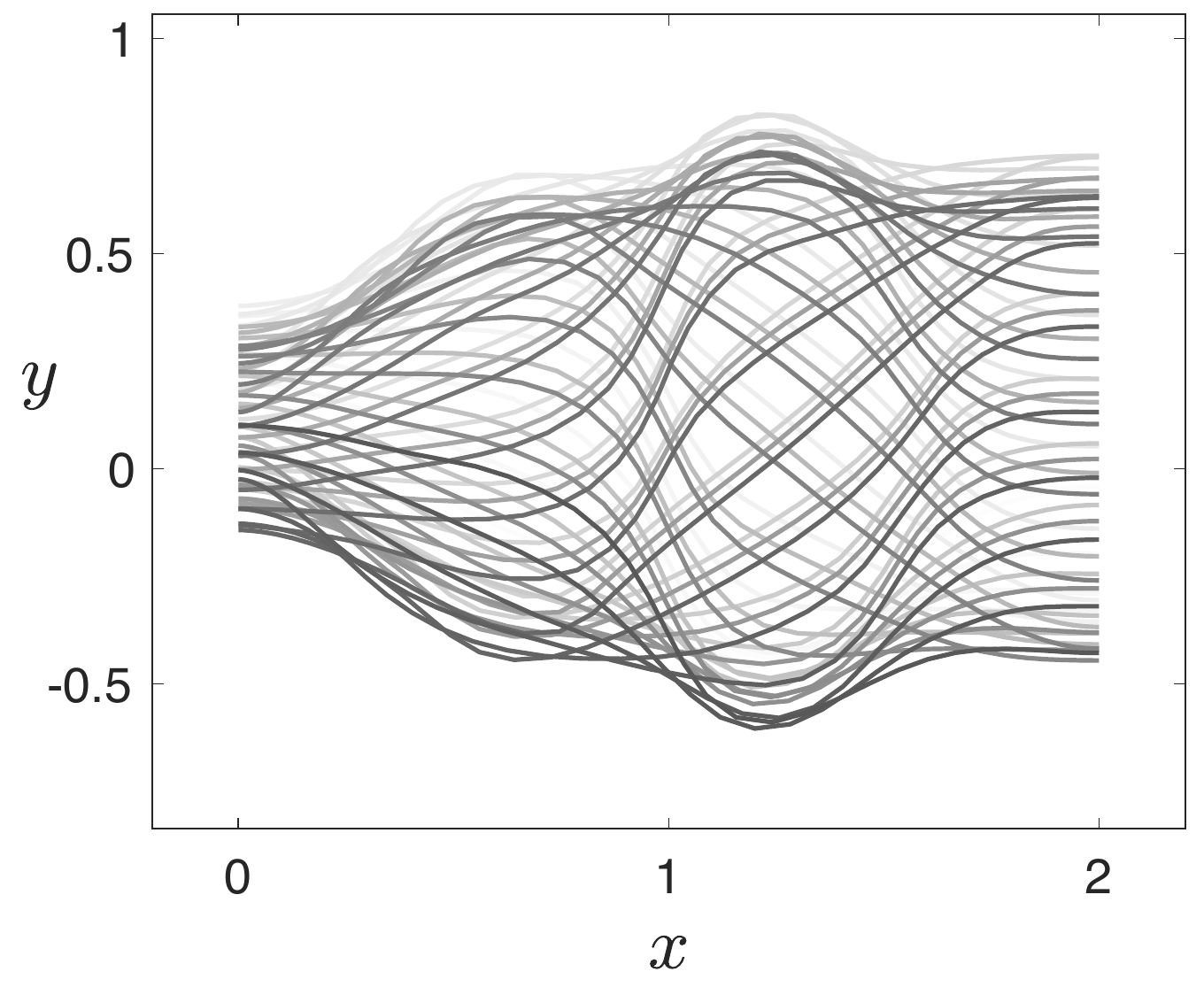}
		\put(-7,98){(a)}
	\end{overpic}\hspace{1.5cm}
	\begin{overpic}[width=.27\textwidth,tics=10]{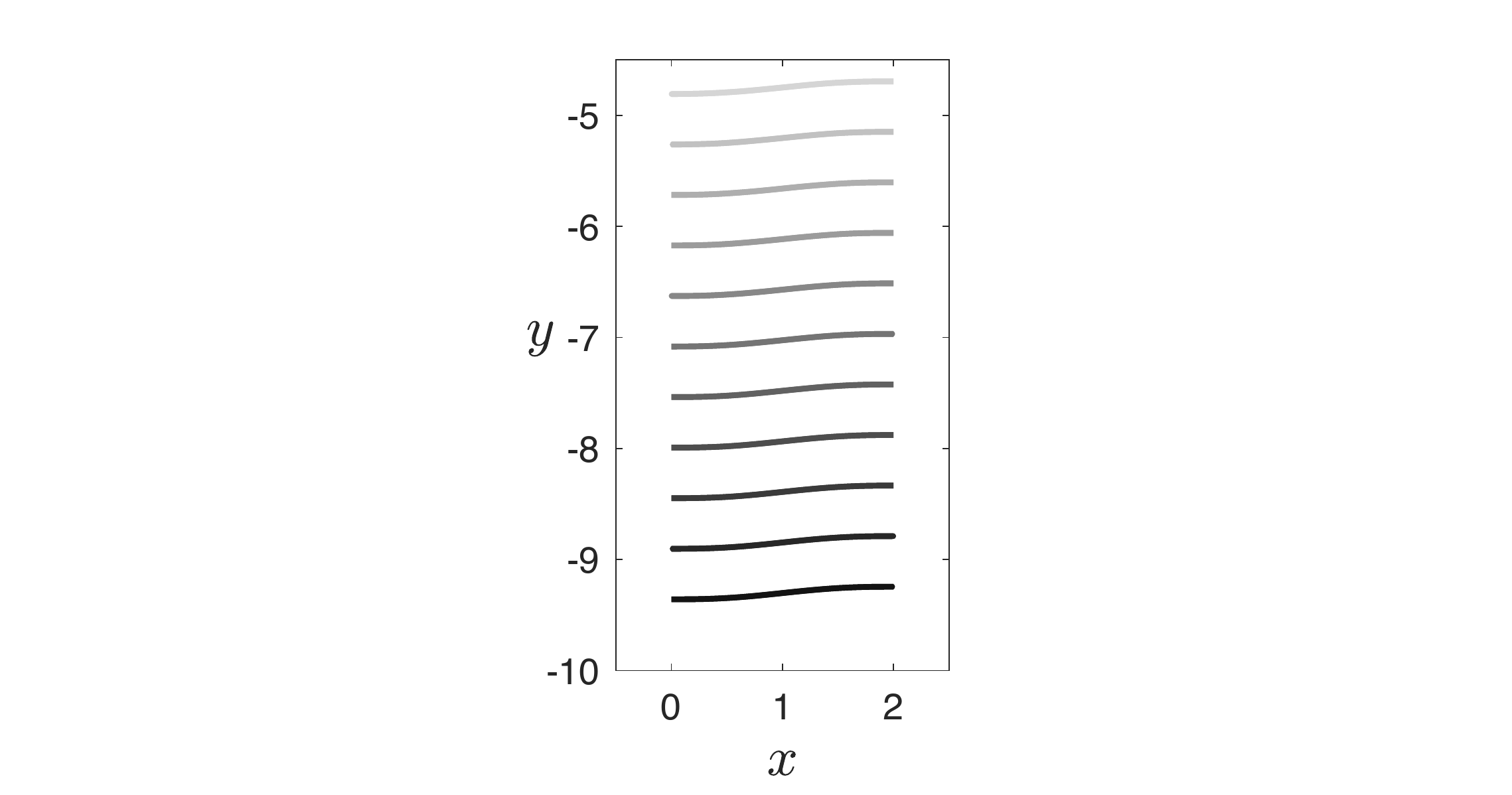}
	\put(-7,95){(b)}
			\put(18,18){Increasing time}
		\linethickness{0.8pt}
		\put(35,98){\vector(0,-1){76}}
	\end{overpic}
	\caption{Membrane snapshots at the large-amplitude regime, for (a) $R_1=10^{0}$ and $R_3=10^{0.5}$ and (b) $R_1=10^{0.5}$ and $R_3=10^{2.5}$. The shading of the membrane indicates the different phases in the motion, varying from gray at earlier times to black at current times.\label{fig:FreeExamples}}
\end{figure}

Examples of vortex wakes in the free-free case are shown in figure \ref{fig:vortex2DFreeFree}. The wakes have oscillatory patterns like those in the fixed-free case (figure \ref{fig:vortex2DFixedFree}). Here, however, the membranes' translational
motion leads to more complexity in the wakes' spatial configurations. Fewer of
these cases resemble a von K\'{a}rm\'{a}n vortex street than those in the fixed-free case. For example, the membrane with $R_1=10^{0}$ and $R_3=10^{0.5}$ in figure~\ref{fig:snapshotsFreeFree} oscillates
almost periodically in the $y$ direction. The motion
is shown enlarged in figure \ref{fig:FreeExamples}(a). The corresponding
vortex wake, shown in figure \ref{fig:vortex2DFreeFree} (middle column, second from bottom) is more complex
than a von K\'{a}rm\'{a}n vortex street.
At the upper right of figure \ref{fig:vortex2DFreeFree} are approximately straight-line wakes, corresponding to membranes that
translate steadily with a constant shape, 
e.g.\ the enlarged example in figure~\ref{fig:FreeExamples}(b).
Here the vortex wakes have zero strength density in the large-time limit, and so they
translate steadily downstream without any self-induced undulatory motion or roll-up.


\begin{table}[H]
	\caption{Comparison of frequencies in the small-amplitude and large-amplitude regimes with $R_3=10^{1.5}$ for three pairs of $(R_1,T_0)$ at the flutter and divergence region for fixed-free and free-free membranes. \label{tab:freq}}
	\centering\vspace{.1cm}
	\begin{tabular}{llcc}
		\hline\noalign{\smallskip}
		& $(R_1,T_0)$ & Small-amplitude frequency & Large-amplitude frequency\\
		\noalign{\smallskip}\hline\noalign{\smallskip}
		&$(10^0,10^{-0.5})$  & 0.1717 & 0.1313\\
		Fixed-free & $(10^{0.5},10^{-0.5})$  & 0.0970& 0.1240\\
		&	$(10^{1.5},10^{0})$  & 0.0362 & 0.0543 \\
		\noalign{\smallskip}\hline\noalign{\smallskip}
		Free-free &$(10^0,10^{-0.5})$  &  0.1607 &  0.1303\\
		&	$(10^{1},10^{0})$  & 0.0241 & 0.0542\\
		\noalign{\smallskip}\hline
	\end{tabular}
\end{table}

Our final results are a brief comparison of membrane frequencies in the small-amplitude exponential growth regime, the focus of previous membrane flutter studies, and the large-amplitude steady-state regime.
The top three rows of table~\ref{tab:freq} compare the small- and large-amplitude frequencies of three fixed-free membranes, shown in figure~\ref{fig:deflT0fixedFree} (at $R_3=10^{1.5}$),
that become unstable through flutter and divergence and oscillate with a single dominant frequency. The bottom two rows compare the frequencies for two free-free membranes that also oscillate with single dominant frequencies. We note that the frequency may become significantly lower or higher as the membranes transition from small to large amplitude. It is unclear in general if aspects of the large amplitude motion can be inferred from the
shapes and frequencies of the unstable modes
in the linearized, small-amplitude regime.



\section{Conclusions}\label{sec:conclusions}

In this work we have studied
the flutter instability and large amplitude dynamics for thin membranes.
These are made of elastic materials---e.g.\ rubber, textile fabric, or the skin of swimming or flying animals---with
Young's moduli sufficiently small
that stretching provides the primary resistance to fluid forces and bending
resistance is negligible. Previous studies have considered the flutter instability of membranes with fixed ends. We find that all such membranes become unstable by divergence below a critical pretension $T_0$ close to the value identified in previous studies. Surprisingly, we find that all cases that exhibit large (but physically reasonable) deflections converge to states of steady deflection with single humps that are almost fore-aft symmetric, and the
deflections scale as $1/\sqrt{R_3}$, where $R_3$ is the stretching modulus. These deformations are similar to those found with linearized models that assumed steady deflection at a fixed angle of attack \cite{waldman2017camber,tzezana2019thrust}. 

We then considered membranes with the leading edge fixed and the trailing edge free, and found a wide range of unsteady dynamics, somewhat similar to those seen in studies of flapping plates or flags. The critical pretension $T_0$ now depends on the membrane mass density $R_1$. Membranes become unstable with divergence or with a combination of flutter and divergence in some cases near the stability boundary. The large-amplitude dynamics are independent of the pretension except close to the stability boundary, where the dynamics are in some cases more periodic and have smaller amplitudes. The dynamics depend most strongly on the membrane mass density $R_1$. At $R_1 \leq 0.1$, a number of very small amplitude motions with high temporal and spatial frequencies occur. Though these have small amplitudes, they are much larger than the initial perturbation and are the steady state motion after the initial exponential instability has saturated. As $R_1$ increases above 0.1, 
the motions have steadily increasing amplitudes and decreasing spatial and temporal frequencies. They become more regular and periodic at first, then increasingly chaotic and asymmetrical at the largest mass. Here the mean temporal frequency scales as $1/\sqrt{R_1}$.

With both edges free, the membrane motions show two new features---a vertical translational component that may be nearly steady or oscillatory, and a nonzero slope. The combination of the two yields a small angle of attack with respect to the oncoming flow. The translational motion may be steady, periodic, or chaotic, and switch among these states with small changes of parameters. 
Superposed on the translational motions with nonzero slope are
modes with oscillatory spatial and temporal features, similar to those in the fixed-free case in how they vary with $T_0$, $R_1$, 
and $R_3$.

Membrane (as opposed to beam/plate) flutter with free ends has barely been explored. One application is to energy harvesting by membranes mounted
on tensegrity structures (networks of rigid rods and elastic fibers) and placed in fluid flows \cite{sunny2014optimal,yang2016modeling}. In such cases the membrane ends have some degrees of freedom akin to the free-end boundary conditions we have used. Extensional deformations may be used in conjunction with (\cite{drachinsky2016limit,chatterjee2018aeroelastic}) or as an alternative to bending-dominated deformations for energy harvesting, e.g.\
the flutter of piezoelectric beams and bilayers \cite{erturk2010energy,giacomello2011underwater,doare2011piezoelectric,porfiri2013energy,kim2013flapping,wang2016stability,orrego2017harvesting}.



\vspace{0.3in}
\noindent {\large Acknowledgments}\\
\noindent We acknowledge support from a Rackham International
Student Fellowship (University of Michigan) to C.M.


\appendix
\section{Pressure jump equation}\label{app:pressureJump}

In this appendix we derive the equation for the pressure jump $[p](\alpha,t)$ across the membrane, given by~\eqref{eq:pressureAlpha}, as in~\cite{alben2012attraction} but for an extensible body. We use vector notation instead of complex notation.

The Euler momentum equation given by 
\begin{equation}\label{euler}
\partial_t \boldsymbol{u}(\boldsymbol{x},t)+ \boldsymbol{u}(\boldsymbol{x},t)\cdot\nabla\boldsymbol{u}(\boldsymbol{x},t)=-\nabla p(\boldsymbol{x},t),
\end{equation}
determines the velocity of the fluid flow $\boldsymbol{u}(\boldsymbol{x},t)$ at a point $\boldsymbol{x}$ in the fluid. 
We want to calculate the fluid pressure at a point in the fluid that is adjacent to and follows a material point $\boldsymbol{X}(\alpha,t)$ on the membrane. The rate of change of fluid velocity at such a point is 
\begin{equation}\label{material}
\frac{d}{dt}\boldsymbol{u}(\boldsymbol{X}(\alpha,t),t)=\partial_t\boldsymbol{u}(\boldsymbol{x},t)|_{\boldsymbol{x}=\boldsymbol{X}(\alpha,t)}+\left(\partial_t\boldsymbol{X}(\alpha,t)\cdot\nabla\right)\boldsymbol{u}(\boldsymbol{x},t)|_{\boldsymbol{x}=\boldsymbol{X}(\alpha,t)}.
\end{equation}
We replace the first term in (\ref{euler})
using the same term in (\ref{material}) (the first term on the right hand side). This yields the pressure gradient at a point that moves with $\boldsymbol{X}(\alpha,t)$.
Since the fluid velocity is discontinuous across
the membrane, we actually need to do this separately for points that tend toward $\boldsymbol{X}(\alpha,t)$ from each side of the
membrane. We obtain
\begin{equation}\label{bothSidespm}
\frac{d}{dt}\boldsymbol{u}(\boldsymbol{X}(\alpha,t),t)^{\pm}+\left((\boldsymbol{u}(\boldsymbol{x},t)-\partial_t\boldsymbol{X})\cdot\nabla\boldsymbol{u}(\boldsymbol{x},t)\bigg|_{\boldsymbol{x}=\boldsymbol{X}(\alpha,t)}\right)^{\pm}=-(\nabla p(\boldsymbol{x},t)|_{\boldsymbol{x}=\boldsymbol{X}(\alpha,t)})^{\pm},
\end{equation}
using $+$ for the side toward which the membrane normal $\boldsymbol{\hat{\mathbf{n}}}$ is directed and $-$ for the other side.

Next, we decompose the fluid velocity into components tangential and normal to the membrane. The normal component matches that of the membrane, $\nu$ in~\eqref{eq:tau}.
The tangential component of the fluid velocity may be written in terms of its jump across the membrane,  the same as the vortex sheet strength $\gamma$~\cite{saffman1992vortex},
\begin{equation}
\left(\boldsymbol{u}^+-\boldsymbol{u}^-\right)\cdot \hat{\mathbf{s}}=-\gamma,
\end{equation}
and the average of the tangential components of the fluid velocity on the two sides of the membrane, denoted $\mu$.
Combining the tangential and normal components we have
\begin{equation}\label{fluidvel}
\boldsymbol{u}^{\pm}=\left(\mu\mp\frac{\gamma}{2}\right)\boldsymbol{\hat{\mathbf{s}}}+\nu\boldsymbol{\hat{\mathbf{n}}}.
\end{equation} 



We take the difference of (\ref{bothSidespm}) on the $+$ and $-$ sides:
\begin{align}\label{diff}
\frac{d}{dt}&\left(\boldsymbol{u}^+(\boldsymbol{X}(\alpha,t),t)-\boldsymbol{u}^-(\boldsymbol{X}(\alpha,t),t)\right)
+\left((\boldsymbol{u}^+(\boldsymbol{x},t)-\partial_t\boldsymbol{X})\cdot\nabla\boldsymbol{u}^+(\boldsymbol{x},t)\bigg|_{\boldsymbol{x}=\boldsymbol{X}(\alpha,t)}\right)\nonumber\\&-  
\left((\boldsymbol{u}^-(\boldsymbol{x},t)-\partial_t\boldsymbol{X})\cdot\nabla\boldsymbol{u}^-(\boldsymbol{x},t)\bigg|_{\boldsymbol{x}=\boldsymbol{X}(\alpha,t)}\right)
=-(\nabla p(\boldsymbol{x},t)^+-\nabla p(\boldsymbol{x},t)^-) |_{\boldsymbol{x}=\boldsymbol{X}(\alpha,t)}.
\end{align}
We then take the tangential component of (\ref{diff}),
term by term. Using (\ref{fluidvel}),
\begin{align}\label{dtdiff}
\hat{\mathbf{s}}\cdot\frac{d}{dt}(\boldsymbol{u}^+(\boldsymbol{X}(\alpha,t),t)-\boldsymbol{u}^-(\boldsymbol{X}(\alpha,t),t)) = \hat{\mathbf{s}}\cdot \partial_t(-\gamma(\alpha,t) \hat{\mathbf{s}}(\alpha,t)) =
-\partial_t\gamma(\alpha,t).
\end{align}
Using
\begin{equation}\label{membranevel}
\partial_t\boldsymbol{X}=\tau\boldsymbol{\hat{\mathbf{s}}}+\nu\boldsymbol{\hat{\mathbf{n}}},
\end{equation}
and (\ref{fluidvel}), 
\begin{align}\label{advpm}
   \hat{\mathbf{s}}\cdot \left[(\boldsymbol{u}^\pm(\boldsymbol{x},t)-\partial_t\boldsymbol{X})\cdot\nabla\boldsymbol{u}^\pm(\boldsymbol{x},t)\right]\bigg|_{\boldsymbol{x}=\boldsymbol{X}(\alpha,t)} = \left(\mu \mp \frac{\gamma}{2} -\tau\right)\left[\partial_s\left(\mu\mp\frac{\gamma}{2}\right) -\nu\kappa\right].
\end{align}
The difference of the $+$ and $-$ terms on the right hand side of (\ref{advpm}) is
\begin{align}\label{advdiff}
-(\mu - \tau)\partial_s\gamma -\gamma(\partial_s\mu-\nu \kappa).
\end{align}
The tangential component of the right hand side
of (\ref{diff}) is
\begin{align}\label{pdiff}
-\partial_s[p]_-^+(\boldsymbol{x},t)|_{\boldsymbol{x}=\boldsymbol{X}(\alpha,t)}.
\end{align}
Combining (\ref{dtdiff}), (\ref{advdiff}), and
(\ref{pdiff}),
the tangential component of (\ref{diff}) is
\begin{equation}
\partial_t\gamma+(\mu - \tau)\partial_s\gamma +\gamma(\partial_s\mu-\nu \kappa)=\partial_s[p]_-^+,
\end{equation}
or using $\alpha$-derivatives,
\begin{equation}
\partial_\alpha s\partial_t\gamma+(\mu - \tau)\partial_\alpha\gamma +\gamma(\partial_\alpha\mu-\partial_\alpha s\nu \kappa)=\partial_\alpha[p]_-^+.
\end{equation}

\section{Membrane frequencies in the free-free case \label{freefreefreq}}

\begin{figure}[H]
	\centering
	\includegraphics[width=.95\textwidth]{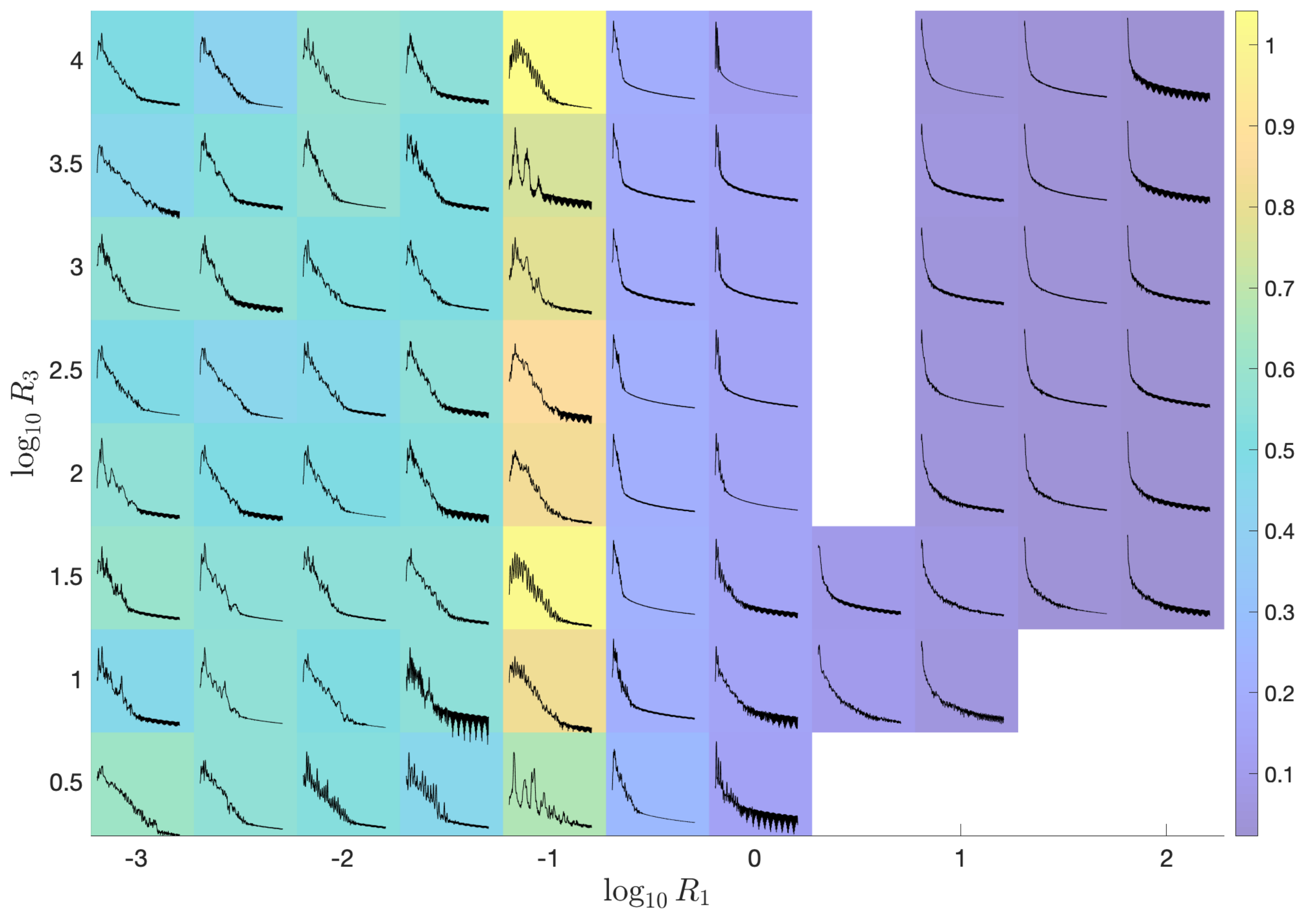}
	\caption{Surface plot of the mean frequency computed from the time series of the circulation, once the membranes have entered the large-amplitude regime, with $T_0=10^{-2}$. The corresponding power spectra for each of the membranes are also shown on the surface plot. The data in the right bottom corner are obtained for a shorter time and so, we neglect the computational results for those values of $R_1$ and $R_3$.}\label{fig:surfMeanFreqFreeFree}
\end{figure}

Since we have determined where in the parameter space the membrane is unstable, we can characterize the large-amplitude dynamics using the mean frequency and study how it depends on $R_1$ and $R_3$. We focus on the region where reliable frequency data can be obtained. As we have done previously, we compute the power spectrum from a plot of the circulation versus time, when the membrane has reached large-amplitude dynamics. In figure~\ref{fig:surfMeanFreqFreeFree}, we see that in general the frequency decreases with increasing $R_1$. For $R_1=10^{0.5}$ and $R_3$ ranging from $10^2$ to $10^4$, the membranes translate steadily (see figure~\ref{fig:deflectionFreeFree}) and the wake
circulation tends to zero at large times. Therefore the power spectra for those cases are zero and we omit them.

\bibliographystyle{unsrt}  
\bibliography{biblio.bib} 

\end{document}